\documentclass[letterpaper]{article} 
\usepackage{aaai23}  
\usepackage{times}  
\usepackage{helvet}  
\usepackage{courier}  
\usepackage[hyphens]{url}  
\usepackage{graphicx} 
\def\UrlFont{\rm}  
\usepackage{natbib}  
\usepackage{caption} 
\usepackage{algorithm}
\usepackage{algorithmic}

\usepackage{newfloat}
\usepackage{listings}
\DeclareCaptionStyle{ruled}{labelfont=normalfont,labelsep=colon,strut=off} 
\floatstyle{ruled}
\newfloat{listing}{tb}{lst}{}
\floatname{listing}{Listing}
\usepackage[utf8]{inputenc} 

\usepackage{booktabs}       
\usepackage{amsfonts}       
\usepackage{nicefrac}       
\usepackage{microtype}      
\usepackage{xcolor}         
\usepackage{pgfplots}
\pgfplotsset{compat=1.15}
\usepackage{mathrsfs}
\usetikzlibrary{arrows}

\usepackage{amsmath,amsfonts,bm}

\def\1{\bm{1}}

\def\rvc{{\mathbf{c}}}

\def\rvh{{\mathbf{h}}}

\def\rvw{{\mathbf{w}}}

\def\rvx{{\mathbf{x}}}

\DeclareMathAlphabet{\mathsfit}{\encodingdefault}{\sfdefault}{m}{sl}
\SetMathAlphabet{\mathsfit}{bold}{\encodingdefault}{\sfdefault}{bx}{n}

\def\gB{{\mathcal{B}}}

\def\gT{{\mathcal{T}}}

\DeclareMathOperator*{\argmax}{arg\,max}

\usepackage{bm}
\usepackage{booktabs}       
\usepackage{amsthm}
\usepackage{nicefrac}       
\usepackage{microtype}      
\usepackage{xcolor}         
\usepackage{xspace}
\usepackage{subfig}
\usepackage{multirow}

\newcommand{\pin}{d}
\newcommand{\ns}{\textsf{ns-3}\xspace}
\newcommand{\pantheon}{\textsf{Pantheon}\xspace}

\newcommand{\numtraces}{\ensuremath{N}}
\newcommand{\numwindows}{\ensuremath{N_w}}

\newcommand{\tracelen}{\ensuremath{T}}
 
\newcommand{\bufferdesc}{Recurrent Buffering Unit}
\newcommand{\buffer}{\textsf{RBU}}
\newcommand{\ThetaWin}{\ensuremath{\Theta_{\text{window}}}}
\newcommand{\ThetaRBU}{\ensuremath{\Theta_{\text{\buffer}}}}
\newcommand{\propd}{\ensuremath{d_{\text{prop}}}}
\newcommand{\transd}{\ensuremath{d_{\text{trans}}}}
\newcommand{\draind}{\ensuremath{d_{\text{trans}}}}
\newcommand{\netpath}{\ensuremath{S \overset{\mathcal{N}}{\leadsto} R}}
\newcommand{\queued}{\ensuremath{d}}
\newcommand{\dqueue}{\ensuremath{d_\text{queue}}}
\newcommand{\RNNConstant}{\textsf{$\text{LSTM}_{\text{win}}$}}
\newcommand{\RNNSampled}{\textsf{$\text{LSTM}_{\text{pkt}}$}}
\newcommand{\RNNConstrained}{\textsf{$\text{LSTM}_{\text{pkt,FIFO}}$}}
\newcommand{\iBoxNet}{\textsf{iBoxNet}}
\newcommand{\Transformer}{\text{Transformer}}
\usepackage{soul}

\newtheorem{proposition}{Proposition}
\newtheorem{definition}{Definition}

\title{Simulating Network Paths with \bufferdesc{}s}
\author{%
  Divyam Anshumaan \equalcontrib, \textsuperscript{\rm 1}
  Sriram Balasubramanian \equalcontrib ,\textsuperscript{\rm 1, \rm 2} \thanks{Work partially done as a Research Fellow at Microsoft Research India} 
  Shubham Tiwari \textsuperscript{\rm 1} \\
  Nagarajan Natarajan ,\textsuperscript{\rm 1} 
  Sundararajan Sellamanickam, \textsuperscript{\rm 1}
  Venkata N. Padmanabhan \textsuperscript{\rm 1}\\
 }

\affiliations{
    \textsuperscript{\rm 1}Microsoft Research India\\
    \textsuperscript{\rm 2}University of Maryland, College Park\\
    t-danshumaan@microsoft.com,
    sriramb@cs.umd.edu,
    t-shutiwari@microsoft.com,
    nagarajn@microsoft.com,
    ssrajan@microsoft.com,
    padmanab@microsoft.com
   
}

\begin{document}
\maketitle

\begin{abstract}

Simulating physical network paths (e.g., Internet) is a cornerstone research problem in the emerging sub-field of AI-for-networking. We seek a model that generates end-to-end packet delay values in response to the time-varying load offered by a sender, which is typically a function of the previously output delays. The problem setting is unique, and renders the state-of-the-art text and time-series generative models inapplicable or ineffective. We formulate an ML problem at the intersection of dynamical systems, sequential decision making, and time-series modeling. We propose a novel grey-box approach to network simulation that embeds the semantics of physical network path in a new RNN-style model called \bufferdesc, providing the interpretability of standard network simulator tools, the power of neural models, the efficiency of SGD-based techniques for learning, and yielding promising results on synthetic and real-world network traces.

\end{abstract}

\section{Introduction}
\label{sec:intro}

Network simulation provides a cost-effective way of developing and evaluating networking applications (e.g. video-conferencing) and protocols. It is a cornerstone research problem, recognized as such by the networking community~\cite{SIGCOMM}, with applications in AI-for-networking~\cite{9351834}.

Network simulation entails delaying or dropping the data packets traversing a sender-receiver network path appropriately. The sender $S$ typically adapts its sending rate continuously based on the feedback in terms of delays or drops gleaned from the packets sent previously. For the simulation to be realistic, the packet delays and drops produced by the simulation mechanism should reflect the target network conditions faithfully, both at the microscopic and the macroscopic levels, so as to recreate application-level metrics such as throughput distribution.

Widely-used network simulation tools such as \ns~\cite{ns} require configuring with a certain network topology,link bandwidth, cross-traffic, etc., typically performed manually by networking domain experts. However, it is extremely challenging to ensure realism in such a manual approach. State-of-the-art (SOTA) data-driven configuration techniques~\cite{yan2018pantheon,ibox} try to mitigate this challenge, but (a) they do not accommodate real-world network behaviors like packet reordering, and (b) scale poorly as they rely on black-box optimization~(Section~\ref{sec:setup}). 

In this work, we formulate and study a novel ML problem of simulating a target network path. The goal is to respond to the sending protocol's actions with \textit{realistic} delay values for every packet, just like the target network would. Note that the sending protocol is provided as input and it could be very different at test vs train. Formally:  

\begin{definition}[End-to-end network path simulation]
\label{def:problem}
Let $(\Pi, \mathcal{N})$ denote the traces collected using a sending protocol $\Pi$, over a target network path $\mathcal{N}$ between a sender $S$ and a receiver $R$, i.e., \netpath~(e.g., the path between a cloud server and a cellular client in certain locations, during the peak hours of a day). We seek a model $\widehat{\mathcal{N}}$ that simulates $\netpath$ s.t. for a new and previously unseen protocol $\Pi'$, the simulated traces $(\Pi', \widehat{\mathcal{N}})$ ``closely match'' the ground-truth $(\Pi', \mathcal{N})$, that would be obtained if we took the trouble of actually running $\Pi'$ too on the same path \netpath~under identical network conditions. The match is in terms of the metrics that networking applications care about; e.g., the joint delay and throughput distribution.
\end{definition}

This problem poses the following key challenges:

\textbf{(A1) Reactive inputs at test time:} At test time, the decisions made by the protocol (e.g., new sending rate), forming the input to the model~$\widehat{\mathcal{N}}$, 
are a 
\textit{response} to the delays output by the model. So, we cannot expect the entire input sequence to be available ahead of time, unlike in standard predictive~\cite{DeepSSM,salinas2020deepar} or generative modeling settings~\cite{esteban2017realvalued,fu2019time,smith2020conditional}. \\
\textbf{(A2) Unseen test protocols:} At test time, the behaviour of inputs to the model (governed by $\Pi'$) can change drastically from that of the training time (governed by $\Pi$), as it depends on \textit{how} the protocol $\Pi'$ reacts to the (simulated) delays.\\
\textbf{(A3) Non-trivial success metric:} The stochastic nature of the network setting means that the metrics of interest are distributional, e.g., joint delay and throughput distribution. Unlike in sequential decision making, we cannot attach a reward to a given output sequence.

We address this challenging problem (Definition 1) by focusing on modeling the behavior of the target path $\mathcal{N}$, using domain-aware neural models, rather than on the actions of the sender protocol that could change drastically at test time. We develop a (conditional) generative modeling technique inspired by network simulation tools like \ns that mimic the components of the physical network. A key aspect is explicitly modeling the \textit{unobservable} cross-traffic which competes for resources on the same network path $\mathcal{N}$ and so critically influences the observed delays.\\
\textbf{Contributions:} We make three key contributions:\\
\textbf{(1)} Novel ML formulation of end-to-end network path simulation --- has significant applications in the development of networking algorithms~\citep{9351834}.\\
\textbf{(2)} A grey-box approach to network simulation that embeds the semantics of physical network path in a new RNN-style model called \bufferdesc~or~\buffer~(Section~\ref{sec:bufferunit}) --- provides the interpretability of simulator tools and the expressive power of neural models.\\
\textbf{(3)} Efficient and practical solution --- scales to sequences of length tens of thousands (leverages domain-specific insights for training in Section~\ref{sec:training}), orders of magnitude more than what the SOTA time-series GAN techniques~\cite{TimeGAN,jarrett2021time} can handle, yet produces realistic traces in synthetic and real-world network settings (Section~\ref{sec:exp}).\\

\begin{figure}[h]
     \centering
     \includegraphics[scale=0.32]{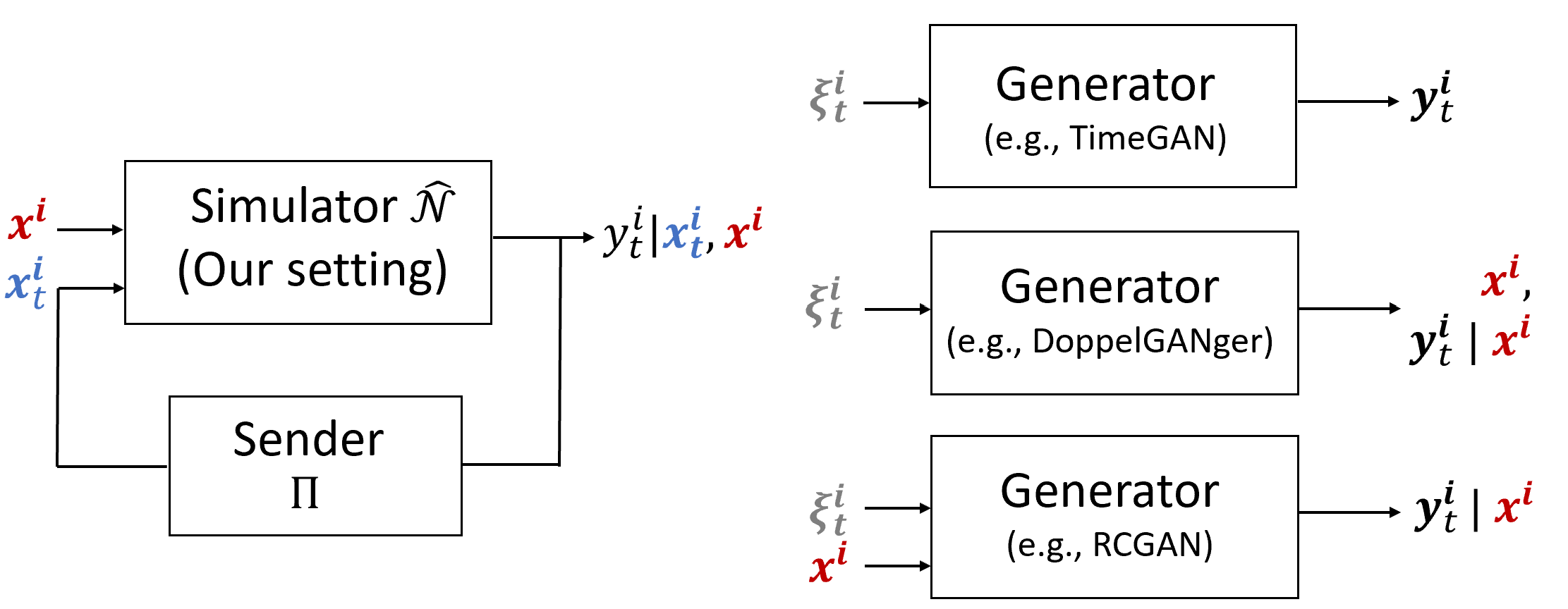}\\
    \caption{Our simulation (left) vs standard generative modeling settings (right); subscript $\cdot_{t}$ denotes time $t$ (absence denotes static feature), and superscript $\cdot^{i}$ denotes series $i$.}
\label{fig:models}
\end{figure}

\textbf{Related Work:} We highlight the relevant ML work here (and revisit some of these in Section~\ref{sec:setup}).

\textit{Generative models for time-series:} SOTA techniques for generating time-series data use RNNs with a GAN-like objective~\cite{esteban2017realvalued,lin2020using,xu2020cot,TimeGAN} or imitation learning~\cite{jarrett2021time}. While they account for longer-range temporal dynamics, error compounding, conditioning on static meta-data, they do not handle \textbf{(A1)}, or scale to very long sequences. Also, evaluation metrics like Maximum Mean Discrepancy, discriminative scores, etc. used in these works are secondary to our domain-specific metrics in \textbf{(A3)}.

\textit{Generative models for text:} In the language domain, recent work have used LSTMs~\cite{sutskever2014sequence,sutskever2011generating} or Transformers~\cite{radford2018improving,radford2019language} to complete or generate sequences, given a context. Indeed the GPT-class models have shown impressive performance in language understanding and text generation tasks, leveraging the self-attention idea in the decoder to capture temporal and positional dependencies while eschewing recurrences of RNNs. However, as with recurrent networks, scaling to very long sequences and obeying domain-specific constraints continue to persist with Transformers, as we observe in our evaluation (Section~\ref{sec:exp}). 

\textit{Sequential decision making/RL:} Our problem has the flavor of sequential decision making in~\textbf{(A1)}. RL formulations applied to such problems~\cite{levine2020offline,ranzatosequence} focus on maximizing expected rewards over multiple trials, which doesn't apply to our setting as stated in~\textbf{(A3)}. Frameworks like imitation learning are also infeasible because of the lack of interactive access to the target $\mathcal{N}$.

\section{Problem Setup, Background, \& Challenges}
\label{sec:setup}
A network trace, collected using a sender $S$ (e.g., file transfer, video call) over a physical network $\mathcal{N}$, is packet-level time-series of measurements $(\rvx_t, y_t), t=0,1,\dots$, where $\rvx_t \in \mathbb{R}^\pin$ denotes the ``input features'' for packet $t$ (e.g., inter-packet spacing $s_t$, packet sizes) characterizing the load offered to $\mathcal{N}$ by $S$; and $y_t \in \mathbb{R}_{> 0}$ the end-to-end delay experienced by packet $t$, with the convention $y_t = \infty$ when packet $t$ was dropped and so never delivered to the destination. Typically, $S$ runs a protocol $\Pi$, e.g., TCP Cubic~\cite{TCP-Cubic-Ha08}, for adapting the sending rate based on the feedback it gets in terms of delays and drops experienced by the preceding packets. We denote a set of such traces by $(\Pi, \mathcal{N})$. Note that $\mathcal{N}$ is a complex black-box system, and we treat it as such. In addition to the \textbf{packet-level} features $\rvx_t$, we also use 3 \textbf{static features}, denoted by $\rvx$ (dropping the subscript $\cdot_{t}$), to model $\widehat{\mathcal{N}}$: 1) the minimum delay or $y_{\text{min}}$ (approximating the end-to-end network propagation delay), 2) the maximum delay or $y_{\text{max}}$, and 3) the 95th percentile throughput (approximating the bottleneck link bandwidth). 

\textbf{Setup:} As in Definition~\ref{def:problem}, we seek a model $\widehat{\mathcal{N}}$, using $(\Pi,\mathcal{N})$ for training, that helps produce realistic end-to-end delays, like the actual physical network path $\mathcal{N}$ between a sender $S$ and a receiver $R$ would, even for a new sender protocol, $\Pi'$, at test time. $\widehat{\mathcal{N}}$ can be deployed for evaluating new protocols, which may be disruptive or infeasible to perform on the target $\mathcal{N}$. The ``goodness'' of the model $\widehat{\mathcal{N}}$ is determined by how well the application metrics such as the distribution of packet delays and throughput, computed over the simulated traces for the unseen protocol $\Pi'$, i.e., $(\Pi', \widehat{\mathcal{N}})$, match the ground-truth $(\Pi', \mathcal{N})$. While obtaining the ground-truth is challenging in general, it is feasible in controlled settings to enable comparison (described in Section~\ref{sec:exp}).

\textbf{How network simulation is done today:} The widely-used solution for network simulation is to use frameworks like~\ns~\cite{ns} that implement the mechanism of physical network components (like links, buffers, end-points) in software. But, it is very challenging to configure them to reflect the target network conditions.
Recent efforts~\cite{yan2018pantheon,ibox} learn the model $\widehat{\mathcal{N}}$ using a simple abstraction of physical network paths (Figure~\ref{fig:buffer}) and domain knowledge-based heuristics. While they show promise for realistic simulation in some settings (Section~\ref{sec:exp}), they (a) are fairly rigid in the type of networks they can model; for instance, they do not accommodate events like link failures
, or packets arriving out of order at the receiver, and (b) rely on black-box optimization techniques (because they work with the~\ns tool directly), e.g., Bayesian Optimization, which makes it challenging to scale.~\citet{ibox} also briefly discuss the challenges of using neural formulations (covered by LSTM-based baselines in Section~\ref{sec:exp}) for network simulation, which we tackle in our work. 

\begin{figure*}[ht!]
    \centering 
    \includegraphics[scale=0.42]{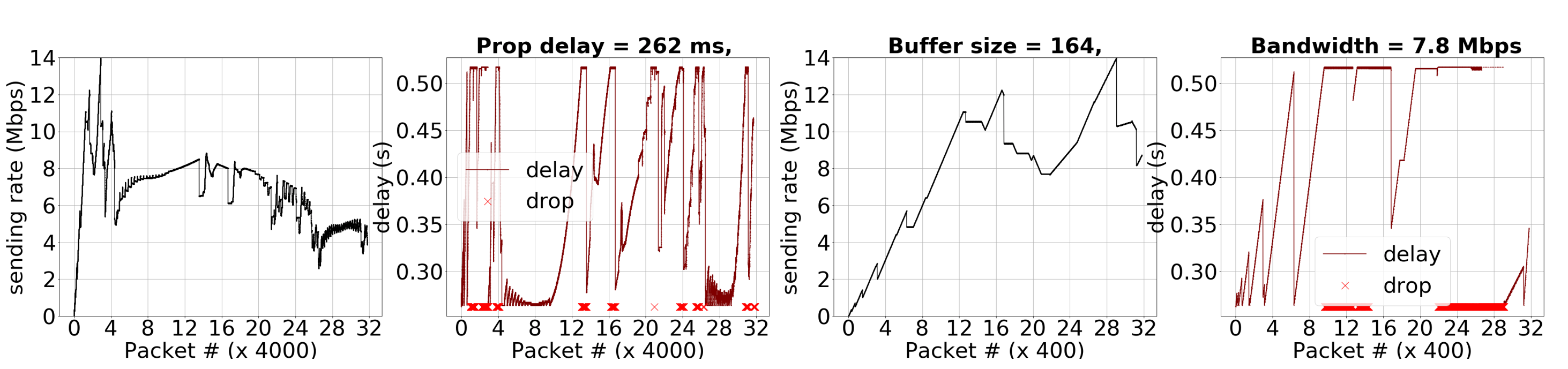}
    \caption{Global (left two) and local (right two, zoomed into the first 10\% packets) characteristics of a typical network trace.}
    \label{fig:gt}
\end{figure*}

\textbf{Inadequacy of domain-agnostic models:}
Consider a typical network trace in Figure~\ref{fig:gt} obtained using $\ns$ tool, configured with a simple topology (in longer version \cite{ibox_ml}), and peak bandwidth of 7.8 Mbps, constituting $\mathcal{N}$. Several observations are in order, from left to right in the Figure. First, the sending rate, regulated by the TCP Cubic protocol at $S$, stabilizes around the peak bandwidth, after a brief initial ``exploration'', characteristic of the protocol. Second, the delays $y_t$ build up to a peak value of around 0.5 seconds, due to the sender behavior \textit{as well as} (unobserved) cross-traffic along $\mathcal{N}$, which together start filling up the bottleneck link buffer. Third, zooming into the start of the trace, the sending rate increases swiftly, and fourth, delay builds up indicating congestion, leading to packet drops once the bottleneck link buffer has been filled up. This example illustrates the \textit{global} (first two plots) and \textit{local} (last two plots) behaviors observed in real-world network traces. These behaviors critically influence the decisions made by the protocol, and in turn the evolution of the network trace, and the application-level metrics. So, learning the model $\widehat{\mathcal{N}}$ entails learning the structure \textit{and} stochasticity in the end-to-end delays $y_t$ conditioned on the inputs seen until and including $\rvx_t$ as well as the previous delays output by $\widehat{\mathcal{N}}$.

At a first glance, this resembles auto-regressive time-series formulations studied in predictive~\cite{borovykh2017conditional,DeepSSM} and generative settings~\cite{sutskever2011generating, DBLP:journals/corr/Graves13}. These techniques factorize the joint $P(y_{1:T} | \cdot)$ into a product of conditionals $\Pi_t P(y_t | \cdot)$, and use RNN-based or Transformer-based~\cite{radford2019language} models to learn a parameterized distribution for the conditional (e.g., multinomial or Gaussian). MLE-based training of these models helps learn $P(y_t|\cdot)$ with a low step-wise loss \textit{in expectation}, but there is no guarantee that \textit{samples} satisfy nuanced dynamics observed in network traces. While such modeling could capture the high-level structure in sequences if carefully trained~\cite{bengio2015scheduled,ranzatosequence}, they fail at capturing the micro-level characteristics (see Section~\ref{sec:exp}), that we articulated using Figure~\ref{fig:gt}.

On the other hand, GAN techniques that directly yield samples~\citep{TimeGAN,xu2020cot,jarrett2021time} are meant for generating synthetic data, which is different from our setting (see Figure~\ref{fig:models}). They do not explicitly model the conditional dynamics~\textbf{(A1)}, and scale poorly with the sequence length --- training TimeGAN~\cite{TimeGAN} to synthesize sequences of length 600 takes over a week on a V100 GPU, with their TensorFlow code. In contrast, network traces are 50x -- 100x longer.

It is quite unclear whether domain-agnostic neural models can capture the fine-grained behaviors in network traces. In the next section, we show how we address the challenges by priming the neural model with domain knowledge in the form of queuing dynamics of real network paths.

\section{Proposed Model: \bufferdesc}
\label{sec:bufferunit}
It seems unlikely, a priori, that the problem posed in Definition~\ref{def:problem}, in the face of challenges \textbf{(A1)--(A3)}, can be solved satisfactorily, even under some assumptions on the sender protocols. The promise comes from a growing body of research underscoring the importance of incorporating the knowledge of physical systems and processes in neural models~\cite{li2020fourier,xu2021conditionally,beucler2021enforcing}. Especially, to tackle \textbf{(A2)}, it is imperative that we model the behavior of the target path $\mathcal{N}$, rather than the network responses to the (observed) actions of the sender protocol --- which could be very different at test time.
Also, any acceptable model for simulating $\mathcal{N}$, in terms of domain-specific metrics~\textbf{(A3)}, should preserve path dynamics at the level of \textit{consecutive} packets. For instance, we want the delays, $y_t$ and $y_{t+1}$, imposed on packets, $t$ and $t+1$, to ensure that these packets delivered at the receiver $R$ are spaced apart in accordance with the bottleneck bandwidth, i.e., a higher (lower) bandwidth would mean a shorter (longer) inter-packet spacing at $R$---otherwise, packets $t$ and $t+1$ being delivered arbitrarily close to each other in time would imply impossibly high available bandwidth for $S$. 

We appeal to how network simulation tools preserve path behaviors and physical constraints \textit{by construction}, i.e., by implementing, in code, the semantics of physical network path composed of links, buffers, and nodes. As we saw in Section~\ref{sec:setup}, the key difficulty in working with such tools is to appropriately configure them. Our first technical contribution is, in essence, to turn the (discrete) simulator tool into a \textit{learnable} model via deriving an end-to-end differentiable formulation.

\begin{figure}[ht]
     \centering
     \includegraphics[scale=0.35]{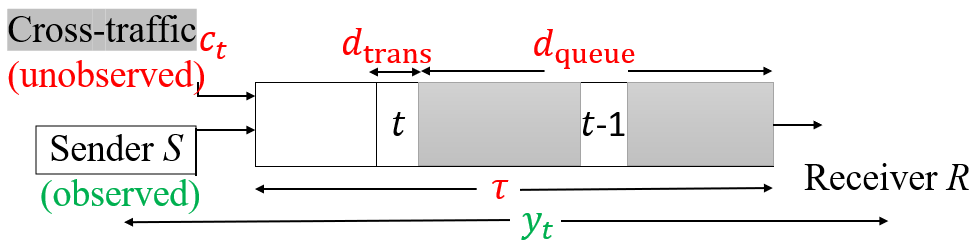}
    \caption{Abstraction of a physical network path.}
\label{fig:buffer}
\end{figure}

Consider an abstraction of the physical path $S \overset{\mathcal{N}}{\leadsto} R$ in Figure~\ref{fig:buffer}. For clarity, we consider a single bottleneck link (where the path is most constrained in terms of bandwidth) of unknown bandwidth $B$ and a FIFO (First-in First-out) queue of unknown buffer size $\tau$, as in~\citep{yan2018pantheon,ibox}. Later in the section, we extend the ideas to multi-path networks, which, among other things, allows us to accommodate phenomena such as packet re-ordering.

The end-to-end delay, $y_t$, suffered by the packet $t$ along the network path in Figure~\ref{fig:buffer} admits a nice structure, comprising (1) the end-to-end propagation delay of $\netpath$, $\propd$, arising from the speed of light, (2) the ``transmission delay'', $\transd \propto 1/B$, or the time taken to transmit a packet onto the network link, and (3) the ``queuing delay" $\dqueue$, or the time spent by packet $t$ in the buffer, waiting for its turn to be transmitted. In other words, $y_t = \propd + (\transd + \dqueue).$

In the rest of the discussion, we define $\transd + \dqueue$ to be the ``bottleneck delay'',  $\queued_t$, for packet $t$, with the subscript $\cdot_{t}$, as the transmission delay and the queuing delay would, in general, vary from packet to packet depending on the packet size and the length of the queue encountered by a packet.

The key component of stochasticity affecting $\queued_t$ are the (unobserved) competing cross-traffic packets $c_t$ also filling the buffer, marked by shaded regions in the figure. 

So, the bottleneck link abstraction of network paths is specified by $\propd, \draind, \tau$, and the dynamic cross-traffic $c_t$.
\paragraph{Modeling parameters}$\propd, \draind, \tau$: While it is possible to estimate these parameters from offline traces using simple heuristics in some cases, such estimates can be grossly inaccurate for real-world traces. So, we model the 3 parameters, with a bounded sigmoid function, in terms of static features and the packet size~(details in Section~\ref{sec:training}).\\
\textbf{Modeling cross-traffic $c_t$:} Cross-traffic typically arises from other senders whose traffic flows via the same bottleneck link buffer. We model the (random) cross-traffic $c_t \in [0,1]$ as a fraction of the \textit{remaining} buffer space occupied when packet $t$ from $S$ arrives. Noting that cross-traffic could react to changes in the network state just as the sender $S$, we model $c_t$ via a non-linear dynamical system: 
\begin{align}
\label{eqn:CT}
\hspace{-0.25cm}c_t = \sigma(\langle \rvw_c, \rvh_{t-1}\rangle), \ \
\rvh_t = \sigma\big(\langle W_h, \cdot\rangle + \langle U_h, \rvh_{t-1}\rangle\big),
\end{align}

where $\rvh_t$ that models the local state of the network path at time $t$ is a standard RNN with weight matrices $U_h$ and $W_h$; the input $\cdot$ to this RNN comprises packet features $\rvx_t$ and the global state of the path as we will see in Section~\ref{sec:training}.\\%

\textbf{Modeling bottleneck delay $\queued_t$:} We derive $\queued_t$ using the FIFO buffer dynamics. Let $s_t$ denote the delta between the sending timestamps of packets $t$ and $t$-$1$, i.e., the inter-packet spacing at $S$. We exploit the following mutually exclusive conditions, when packet $t$ arrives in the buffer. If there are no other packets from $S$ ahead in the queue, i.e, $s_t \geq \queued_{t-1}$, then $\queued_t$ is proportional to the cross-traffic in the queue; else, the packet $t$-$1$ has not yet drained, and the additional delay $a_t$ accrued by packet $t$ is $\queued_{t-1}-s_t$. With $c_t$ modeled as in~\eqref{eqn:CT} and $\text{ReLU}(z) = \max(z, 0)$, we have:
\begin{align}
\label{eqn:qt}
\hspace{-0.3cm}a_t = \draind +  \text{ReLU}\big(\queued_{t\text{-}1} - s_t\big)\ , \
\queued_t &= a_t +  c_t \big(\tau - a_{t}\big).
\end{align}

Note that when $s_t 
\geq\queued_{t-1}$, $\queued_t$ increases with the fraction $c_t$ of cross-traffic occupying the buffer of capacity $\tau$; when $s_t < \queued_{t-1}$, $\queued_t$ increases with the fraction $c_t$ of cross-traffic occupying the buffer of (shrunk) capacity $\tau - a_t$.  

\paragraph{Modeling the output:} The end-to-end delay $y_t$ and the packet drop probability $p_t$ are given by:
\begin{align}
\label{eqn:RBU}
    y_t = \queued_t + \propd, \ \ \text{and} \ \ p_t = \sigma(\queued_t - \tau),
\end{align}
where $\queued_t$ is given by~\eqref{eqn:qt} and $\sigma(\cdot)$ is the sigmoid function. 
\paragraph{\bufferdesc:} The proposed~\bufferdesc~(\buffer) for modeling end-to-end network delays and packet drops (Figure \ref{fig:training}, right) is given by recurrences~\eqref{eqn:CT},~\eqref{eqn:qt},~\eqref{eqn:RBU}.

\begin{proposition}
\label{prop:RBU}
\emph{\buffer} preserves the semantics of single-bottleneck link network path in Figure~\ref{fig:buffer}, when $\sigma$  in~\eqref{eqn:RBU} is the step function. That is, for any two packets originating at $S$ at timestamps $t$ and $t'$, with $t < t'$, they are delivered in order at $R$, i.e., their output delays satisfy $t + y_t < t' + y_{t'}$, or packet $t$ is dropped.  
\end{proposition}
\textbf{Proof:} It suffices to show that $t' + \queued_{t'} > t + d_{t}$. Note that $t' \geq t + s_{t'}$. If we show the inequality for $t' = t + s_{t'}$, when $t'$ is indeed the immediate next packet, then it holds for all future packets. In this case, it reduces to showing $s_{t'} + d_{t'} > d_t$ or $d_{t'} > d_t - s_{t'}$. Now, from~\eqref{eqn:qt}, we have either $\queued_{t'} \geq a_{t'}$ or packet $t$ is dropped, in which case we are done, since $a_t > \tau$ and $c_t \in [0,1]$ together imply $d_t > \tau$. So, $\queued_{t'} \geq a_{t'} = \max(\queued_t - s_{t'}, 0) + \draind > \max(\queued_t - s_{t'}, 0) \geq \queued_t - s_{t'}$, which proves the claim.

\paragraph{Multi-path networks:} In reality, links can fail momentarily over the course of a call, or packets may be randomly routed via different paths and arrive out of order at the receiver. To this end, we consider a generalized multi-path abstraction with a bottleneck link along each path $k$ parameterized by $\draind^{(k)}$ and $\tau^{(k)}$. Consider a packet that enters queue $k$. To model the dynamics here, we need to keep track of the time elapsed between the immediately preceding packet that entered the same queue $k$ and the current packet. We  denote this quantity by $s_{t}^{(k)}$, and update it using the recurrence: $s_0^{(k)} = 0$ and $s_t^{(k)} = s_{t-1}^{(k)} + s_t$. With $c_t$ as in~\eqref{eqn:CT}, the bottleneck delay 
for the queue $k$ that packet $t$ enters is given by recurrences analogous to~\eqref{eqn:qt}:
\begin{eqnarray}
\label{eqn:multiqueue}
    a_t^{(k)} = \draind^{(k)} +  \text{ReLU}\big(\queued_{t-1}^{(k)} - s_{t}^{(k)}\big)\ \ ,\\     
    \queued_t^{(k)} = a_t^{(k)} +  c_t \big(\tau^{(k)} - a_{t}^{(k)}\big) \ 
\end{eqnarray}
For all other queues $k' \neq k$ at time $t$: $a_t^{(k')} = a_{t-1}^{(k')}$, and $\queued_t^{(k')} = \queued_{t-1}^{(k')}$. 
Then, we obtain $y_t$ and $p_t$ from~\eqref{eqn:RBU} with $\queued_t = \queued^{(k)}_t$ and $\tau =  \tau^{(k)}$. Finally, we reset the accumulative elapsed time of the queue $k$, i.e., $s_{t}^{(k)} = 0$, as the current packet has entered queue $k$.



\section{Training and Inference}
\label{sec:training}
\begin{figure}
     \centering
     \includegraphics[scale=0.5]{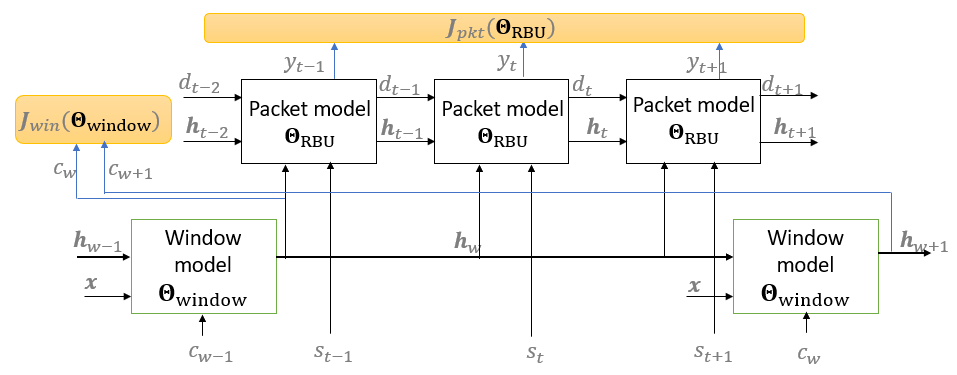}
     \includegraphics[scale=0.5]{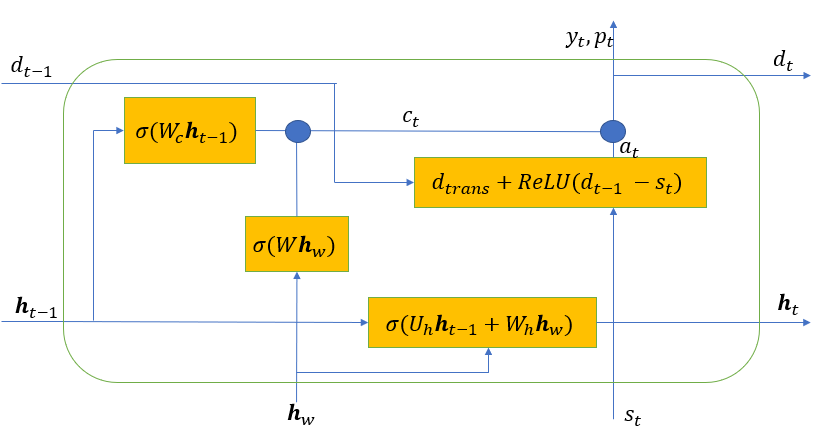}
    \caption{\textbf{Top}: Window-level (LSTM) and packet-level (\buffer{}) models unrolled across time (i.e., packets). The orange boxes indicate training losses. \textbf{Bottom}: \buffer{} cell.} 
\label{fig:training}
\end{figure}
There are three key challenges in learning the \buffer~model, using the traces $(\Pi, \mathcal{N})$.\\
\textbf{(C1) Very long traces.} Traces are extremely long in general (tens of thousands of packets). Trying to jointly learn all the model parameters in the recurrence relations, even in the single bottleneck link case,~\eqref{eqn:CT},~\eqref{eqn:qt} and \eqref{eqn:RBU}, using only the observed end-to-end delays in the traces, may be ill-posed. Working with aggregated or sub-sampled traces is out of question in the simulation setting, unlike in synthetic data generation setting of GANs, because we need to respond at a packet-level to the sender protocol at test time. On the other hand, we could divide the (packet-level) traces into independent chunks that are sufficiently small amenable to efficient training. However, the independence of the chunks means that the global temporal structure of traces (e.g., slow build-up of congestion and surges in cross-traffic) is not captured. We address this challenge using a two-level architecture that preserves both the global and fine-grained characteristics of traces, yet being computationally and sample-efficient. We train a packet-level model within the confines of individual chunks, but the global structure of the traces is integrated via a coarser, window-level model.\\ 
\textbf{(C2) Cross-traffic $c_t$ estimation.} Despite the structure \buffer~model imposes on the delays unlike a vanilla RNN, training the model using standard MLE techniques~\cite{DBLP:journals/corr/Graves13,borovykh2017conditional,salinas2020deepar} does not perform well for our problem, owing to the absence of direct feedback, especially $c_t$~\eqref{eqn:CT}. Using domain knowledge and the global trace structure via the window-level model, we estimate packet-level $c_t$ robustly. \\
\textbf{(C3) Discrete path selection.} In the multi-path scenario, recurrences involve a discrete step of selecting a queue $k$ for packet $t$. This introduces discontinuity in the model at training. In our implementation, we use a smoothed version of recurrences.\\

\textbf{Leveraging global structure:} Consider the step-wise loss (i.e., packet-level), in the spirit of auto-regressive formulations~\cite{DBLP:journals/corr/Graves13,borovykh2017conditional,salinas2020deepar}, to learn the~\buffer~model $\ThetaRBU$:
\begin{equation}
J_{\text{pkt}}(\cdot; \ThetaRBU) := \sum_{i=1}^{\numtraces}\sum_{t=1}^{\tracelen_i} \ell_{\text{pkt}}\big((\hat{y}_{t}^{(i)}, \hat{p}_t^{(i)}), y^{(i)}_t\big),
    \label{eqn:packet}
\end{equation}

where $(\hat{y}_t^{(i)}, \hat{p}_t^{(i)})$ denote the delay and drop probability for packet $t$ in trace $i$ predicted by~\buffer, and $\ell_{\text{pkt}}$ is the loss in~\eqref{eqn:packetloss}. We can apply SGD to minimize~\eqref{eqn:packet}, with standard tricks like chunking, mini-batching, and (truncated) back-propagation through time~\cite{sutskever2013training}. 
But, this performs poorly in our evaluation (Section~\ref{sec:exp}) given extremely long traces.

We devise a two-level architecture (Figure~\ref{fig:training}) that helps mitigate the issues. We use a coarser window-level model to obtain an embedding of the global state of the network path, which then provides crucial feedback on cross-traffic $c_t$ for the packet-level \buffer~model. For the window-level model, we use (2-layer) LSTM, parameterized by $\ThetaWin$ and static trace features $\rvx$ as input, to compute an embedding $\rvh_{w}$ of the path $S \overset{\mathcal{N}}{\leadsto} R$ state. The model operates over fixed-length (100 ms, in experiments, corresponding to the round-trip time on typical network paths when the global state could change), non-overlapping windows:
\begin{equation}
    \rvh_{w} = \text{LSTM}(\rvh_{w-1}, \rvx; \Theta_{\text{window}})\ .
\label{eqn:window}
\end{equation}

\textbf{Estimating cross-traffic:} 
We estimate the \textit{expected} fraction of cross-traffic filling the bottleneck buffer, denoted $c_w \in [0,1]$, using a linear layer over the global path state $\rvh_{w}$ with sigmoid activation. We modify the packet-level $c_t$ in ~\eqref{eqn:CT} to incorporate this information with a hyper-parameter $\gamma \in [0,1]$:

\begin{align}
\label{eqn:CTupdated}
    c_t &= (1-\gamma)\ c_w + \gamma\ \sigma(\langle \rvw_c, \rvh_{t-1}\rangle)\ .
\end{align} 

Furthermore, we obtain a crude estimate of $c_w$ from the training data, by inverting the~\buffer~recurrences to approximate $c_t$; to do so, we use $\gamma = 0$ in~\eqref{eqn:CTupdated} and simple heuristic estimates~\cite{ibox} for $\draind, \propd$ and $\tau$ (discussed in Section~\ref{sec:exp}). We then use the distribution of (discretized) $c_t$ values for packets $t \in $ window $w$, denoted $\tilde{\rvc}_w$, as the ``ground-truth'' for $c_w$, to compute the cross-entropy term:
\begin{equation}
J_{\text{win}}(\cdot; \ThetaWin) := \sum_{i=1}^{\numtraces}\sum_{\text{window }w} \ell_{\text{CE}}\big(c^{(i)}_w, \tilde{\rvc}^{(i)}_w\big) \ .
\label{eqn:windowloss}
\end{equation}
We show in Section~\ref{sec:exp} that even the crude estimate~$\tilde{\rvc}_w$ helps improve the performance of~\buffer~significantly, by providing an effective ``domain-specific regularization'' while training.

\textbf{\buffer~training:} The input to the RNN $\rvh_t$ of the cross-traffic model~\eqref{eqn:CTupdated} comprises $\rvx_t$ and $\rvh_w$ obtained from the window model~\eqref{eqn:window}. The parameters $\propd, \draind, \tau$ must satisfy certain physical constraints; in particular, for a trace with static features $y_{\text{min}}$ and $y_{\text{max}}$, $0 < \propd, \draind \leq y_{\text{min}}$ and $0  < \tau \leq y_{\text{max}} $ by definition. So, we use the bounded sigmoid function for estimating each of the 3 parameters: 
\begin{equation}
g(\rvx) = (b_\rvx - a_\rvx) \sigma(\langle \rvw_{g}, \rvx\rangle) + a_\rvx,
\label{eqn:staticg}
\end{equation}
where $a_\rvx$ and $b_\rvx$ are the lower and upper bounds respectively for the parameter, given $\rvx$. Let $p = \mathbb{1}_{\{y = \infty\}}$ denote the packet drop status in the training data, i.e., $p = 1$ when a packet is dropped (i.e., $y = \infty$), else $p = 0$. We use squared loss for delays and cross-entropy loss $\ell_{\text{CE}}$ for drops:
\begin{equation}
\ell_{\text{pkt}}\big((\hat{y}, \hat{p}), y\big) = p\ \ell_{\text{CE}}(\hat{p}, 1) + (1 - p)(\ell_{\text{CE}}(\hat{p}, 0) + (\hat{y} - y)^2).
\label{eqn:packetloss}
\end{equation}
We set up the optimization problem
 (Figure~\ref{fig:training}):
\begin{equation}
\min_{\ThetaRBU, \ThetaWin} J_{\text{pkt}}(\cdot; \ThetaRBU) + \lambda J_{\text{win}} (\cdot; \ThetaWin),
 \label{eqn:finaleq}
\end{equation}
where $\ThetaRBU$ is the set of RNN weights in~\eqref{eqn:CT}, and the weights for $g$ in~\eqref{eqn:staticg} needed to compute $\propd, \draind, \tau$. We use stochastic gradient-descent to learn the model parameters jointly, with mini-batching, and weight decay on the model parameters.  

\textbf{\buffer~inference:} During simulation, the sender $S$, configured with protocol $\Pi'$, transmits data for the duration of the run (1 minute, in our evaluation). Unbeknownst to $S$, we replace the real network $\mathcal{N}$ with the trained~\buffer~model $\widehat{\mathcal{N}}$. We sample the static features $\rvx$ uniformly from the training data $(\Pi, \mathcal{N})$. We sample $c_w$, needed in~\eqref{eqn:CTupdated}, for $1 \leq w \leq N_w$ ($= 600$, corresponding to 100ms windows over 1 minute), by unrolling the window-level LSTM~\eqref{eqn:window} at once with input $\rvx$. For each packet $t$ that $S$ sends out, we form the features $\rvx_t$ needed as input, together with $\rvx$, for $\widehat{\mathcal{N}}$. We do a forward pass of $\widehat{\mathcal{N}}$ on the input (which takes around 2 ms on a standard GPU for \buffer). The output (delay value or packet drop) from $\widehat{\mathcal{N}}$ is provided as feedback to $S$. Then, $S$ sends out the next packet $t+1$, with inter-packet spacing of $s_{t+1}$, as determined by $\Pi'$ acting on the model feedback, and so on. 


\paragraph{Multipath \buffer~training:} In the multi-path scenario, first note that we can rewrite the set of recurrences~\eqref{eqn:multiqueue} for the links when packet $t$ arrives, using the indicator function $\mathbf{1}_t (k) = 1$, if packet $t$ enters queue $k$ and $\mathbf{1}_t (k') = 0$, for $k' \neq k$. To mitigate~\textbf{(C3)}, we relax this indicator function, during training, with the probability of traversing queue $k$ for packet $t$, denoted by $q_t(k)$. In some cases, we may be able to model $q_t(k)$ using indirect observations in the training traces. For instance, in the two-path case, the fraction $\tilde{q}_w$ of packets sent by $S$ that arrived out-of-order at $R$ in a given time window $w$ (which can be easily computed for any trace) gives an approximation to $q_t(2)$ (or $q_t(1)$ which is $1 - q_t(2)$) for $w$. We estimate $q_w$ from the window-level embedding~\eqref{eqn:window}, just as $c_w$ above, via incorporating a loss term $\ell_{\text{CE}}(q_w, \tilde{q}_w)$ corresponding to~\eqref{eqn:windowloss}. Details of the training and inference procedure for the multi-path scenario are given in the longer version \cite{ibox_ml}. In Section~\ref{sec:reordering_exp}, we demonstrate how this technique helps model multi-path behaviors like packet reordering in real-word network traces.

\begin{table}[h]

\begin{tabular}{p{1cm} p{1.6cm} p{2cm} p{2cm}}
\toprule

 Protocol & Model & WD (Tput, Delay) & WD (P95 Delay) \\ \midrule

 & \iBoxNet & \textbf{0.015} $\pm$ 0.000 &  \textbf{0.000} $\pm$ 0.000 \\
 & \RNNSampled & 0.271 $\pm$ 0.011 &  0.155 $\pm$ 0.001 \\
Cubic & \RNNConstant & 0.164 $\pm$ 0.002 &  0.150 $\pm$ 0.001 \\
 (Train) & \RNNConstrained & 0.214 $\pm$ 0.009 &  0.119 $\pm$ 0.000 \\
 & \Transformer & 0.224 $\pm$ 0.015 & 0.030 $\pm$ 0.003 \\
 & \buffer & 0.032 $\pm$ 0.004 &  0.007 $\pm$ 0.000 \\
\midrule 
 & \iBoxNet & 0.054 $\pm$ 0.000 & 0.088 $\pm$ 0.000\\
 & \RNNSampled & 0.084 $\pm$ 0.000 & 0.112 $\pm$ 0.001  \\
Vegas & \RNNConstant & 0.108 $\pm$ 0.003 & 0.110 $\pm$ 0.001 \\
(Test) & \RNNConstrained & 0.061 $\pm$ 0.000 & 0.091 $\pm$ 0.000 \\
 & \Transformer & 0.079 $\pm$ 0.003 & \textbf{0.007} $\pm$ 0.001 \\
 & \buffer & \textbf{0.041} $\pm$ 0.004 & 0.057 $\pm$ 0.007 \\
\bottomrule 
\end{tabular}
\centering
\begin{tabular}{p{1cm} p{1.6cm} p{2cm} p{2cm}}
\toprule
Protocol & Model & WD (Tput, Delay) & WD (P95 Delay) \\ \midrule
 & \iBoxNet & 0.056 $\pm$ 0.000 & \textbf{0.005} $\pm$ 0.000 \\
 & \RNNSampled & 0.156 $\pm$ 0.002 & 0.154 $\pm$ 0.001 \\
LEDBAT & \RNNConstant & 0.118 $\pm$ 0.019 & 0.149 $\pm$ 0.000 \\
(Test) & \RNNConstrained & 0.106 $\pm$ 0.000 & 0.122 $\pm$ 0.000 \\
 & \Transformer & 0.103 $\pm$ 0.001 & 0.043 $\pm$ 0.005 \\
 & \buffer & \textbf{0.049} $\pm$ 0.000 & 0.008 $\pm$ 0.000 \\
\midrule
 & \iBoxNet & \textbf{0.053} $\pm$ 0.000 & \textbf{0.007} $\pm$ 0.000 \\
 & \RNNSampled & 0.207 $\pm$ 0.009 & 0.152 $\pm$ 0.001 \\
NewReno & \RNNConstant & 0.133 $\pm$ 0.012 & 0.147 $\pm$ 0.001 \\
(Test) & \RNNConstrained & 0.166 $\pm$ 0.007 & 0.124 $\pm$ 0.000 \\
 & \Transformer & 0.165 $\pm$ 0.013 & 0.038 $\pm$ 0.002 \\
 & \buffer & 0.094 $\pm$ 0.018 & 0.024 $\pm$ 0.011 \\
\bottomrule
\end{tabular}
\caption{(mean $\pm$ std. dev) Wasserstein distances (WD), the lower the better, for traces obtained via different models and protocols, on the~\ns~data. The best numbers are in \textbf{bold}.}
\label{tab:WD}
\end{table}

\section{Experiments}
\label{sec:exp}

\textbf{Compared methods:} We compare the~\buffer~model with: (1)~\iBoxNet~\citep{ibox}: a SOTA network simulation approach that uses network domain knowledge to infer parameters from network packet traces, (2)~\RNNConstant: Autoregressive modeling of delays~\citep{sutskever2011generating, DBLP:journals/corr/Graves13} (referred to as ``T-forcing'' in~\cite{TimeGAN,jarrett2021time,xu2020cot}) as the factorized conditional $\Pi_t P(y_t|\cdot)$, implemented with LSTMs trained on windowed traces using $\rvx_t$ as input and (discretized) $y_t$ as output; at inference, we use the $\argmax$ of the output distribution as the delay value for all the packets in the window, (3)~\RNNSampled: Same as~\RNNConstant~during training, but we sample a value from the output distribution independently for each packet in the window at inference, (4) \RNNConstrained: same as~\RNNSampled, but enforces no-packet-reordering constraint (Proposition~\ref{prop:RBU}) while sampling delays, (5) Transformer: We use a GPT (decoder) model~\cite{radford2018improving}. \\
\textbf{Datasets:} We (1) design a synthetic benchmark using \ns as in~\cite{ibox}, consisting of 4200 traces for 4 different TCP protocols, on a variety of cross-traffic patterns and network configurations; and (2) use a subset of traces from a real physical network testbed \pantheon~\cite{yan2018pantheon} for 2 TCP protocols. The \ns~data corresponds to single-path configuration (hence, no packet reordering), while the \pantheon~data includes naturally occurring reordering from real networks. 

In all our experiments, \textit{we use only the TCP Cubic protocol} (dominant on the internet) \textit{traces for training}, and \textit{the other TCP protocols} (Vegas, NewReno, and LEDBAT) \textit{for testing}.\\
\textbf{Implementation:} We implement all the models in PyTorch. The static trace features are normalized to [0,1]. For~\RNNConstant~and~\RNNSampled, we (a) normalize the delays and the sending rates, and (b) use a 2-layer LSTM with 256 hidden units and a fully connected layer with discretized $y_t$ as output (100-dimensional), tuned to maximize mean delay and throughput distribution match, on the training protocol. For~\buffer, we (a) use the same LSTM architecture, to be consistent, for the window-level model in~\eqref{eqn:window}, with discretized $c_w$ in~\eqref{eqn:CTupdated} as output, (b) set $\gamma = 0.1$ in \eqref{eqn:CTupdated} and size of $\bm{h}_t$ in~\eqref{eqn:CT} to 1, which works well across datasets, and (c) use single-bottleneck buffer~\buffer~model (just as the ground-truth) for~\ns, and 2-path~\buffer~model for~\pantheon. For \iBoxNet, we use their official code. Training~\buffer~on the largest dataset (\ns) takes only about 3 minutes per epoch on V100 GPU. \textit{We report mean and std. dev., over 3 independent simulations, for all the metrics.}\\
\textbf{Note:} We give all the key results in this section. For additional details on datasets, implementation, metrics, and for more comprehensive qualitative and quantitative results, we defer the reader to the longer version of our paper~\cite{ibox_ml}.

\subsection{Qualitative evaluation of traces}

\begin{figure*}
    \centering
    \includegraphics[scale=0.4]{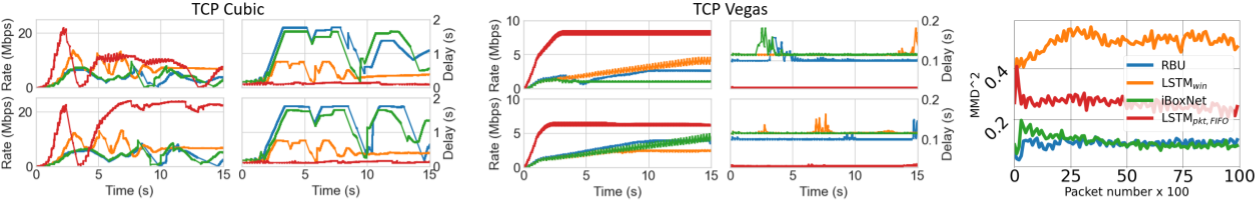}
    \caption{First two plots: (Top row) Ground-truth sending rates, delays for 4 sample TCP Cubic \& Vegas traces; (Bottom row) traces from the \buffer~model trained on Cubic, tested on Cubic and Vegas. Last plot: MMD$^2$ vs chunks for TCP Vegas (\ns~data).}
    \label{fig:rbu-cubic-comparison}
\end{figure*}

In the top row of Figure~\ref{fig:rbu-cubic-comparison}, we show 4 randomly picked ground-truth (GT) traces for Cubic (train) and Vegas (test) protocols from the \ns~dataset. Each trace, shown in a different color for a protocol, consists of a sending rate series and a delay series, trimmed to the first few seconds to show the local behaviors, in separate plots. The bottom row shows (a) for Cubic, traces obtained by running the~\buffer~model with static features obtained from the same 4 GT Cubic traces (to enable direct comparison), and (b) for Vegas, example \buffer~traces that give similar throughput as the 4 GT traces. 

Note how the~\buffer~traces reflect the much lower delays with Vegas vs Cubic, just as in GT. \buffer~is able to achieve such accurate recreation, even though it was trained only on Cubic data.

\begin{figure*}[ht]
    \includegraphics[width=\textwidth]{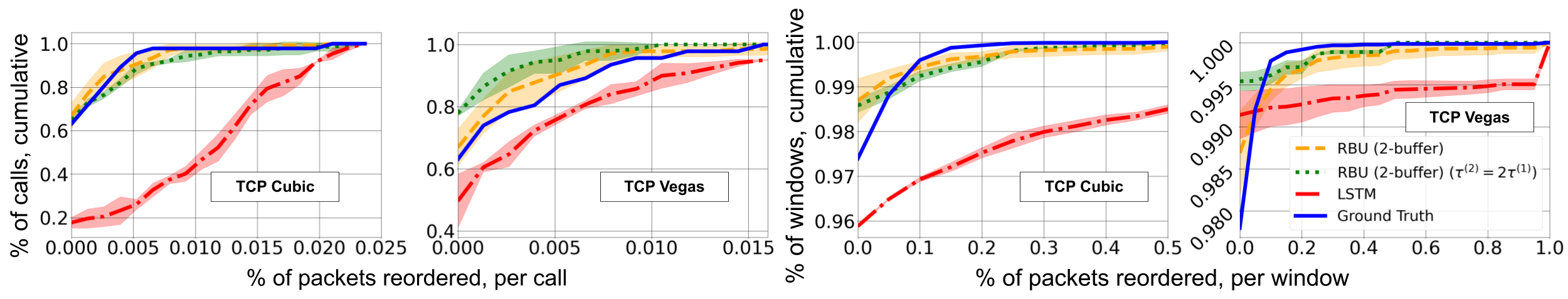}
    \caption{Fraction of packets reordered in calls (first two), windows (last two) for TCP Cubic, Vegas (\pantheon{})}
\label{fig:reordering-plots-cubic-vegas}
\end{figure*}

\subsection{Quantitative evaluation on unseen protocols}
\label{sec:unseen_exp}

We first look at application-level metrics obtained via different models on the~\textbf{\ns~data}. We quantify the distributional match using the standard Wasserstein distance (WD). In Table~\ref{tab:WD}, we present the 2-dimensional WD for the joint throughput, mean delay distribution, and WD for the P95 delay distribution. \buffer~is competitive w.r.t. the SOTA~\iBoxNet~across all the protocols. LSTM-based baselines, on the other hand, fare poorly (as hypothesized in Section~\ref{sec:setup}). The \RNNConstrained~method, where we explicitly constrain the sampled delays to satisfy no re-ordering (as in the GT traces), does improve over the standard variants, but is often worse compared to~\buffer. Notably, we find that the Transformer (GPT) model outperforms the LSTM-based models in many cases (as one would expect), but is not as good as~\buffer. As noted in Section 2, our intuition for this is that (a) it is important to model the delay dynamics carefully, which the Transformers do not, and (b) Transformer models continue to suffer from the pitfalls of LSTM models at inference time. In Table~\ref{tab:WD_pantheon}, we see that~\buffer~performs significantly better than the SOTA~\iBoxNet~on the real-world ~\textbf{\pantheon~data} on both Cubic (train) and Vegas (test) protocols. 

\begin{table*}[ht!]
\centering
\begin{tabular}{l l r r r r}
\toprule
\multirow{2}{*}{Protocol} &
  \multirow{2}{*}{Model} &
  \multicolumn{4}{c}{Wasserstein Distances} \\ 
 &
   &
   2D (Tput, Mean Delay) &
2D (Tput, P95 Delay) &
1D Mean Delay &

1D P95 Delay \\ \toprule
\multirow{4}{*}{Cubic (Train)} &
  iBoxNet &
  \multicolumn{1}{c}{0.150 $\pm$ 0.000} &
  \multicolumn{1}{c}{0.125 $\pm$ 0.000} &
   0.142 $\pm$ 0.000&
  0.116 $\pm$ 0.000 \\ 
 &
  \RNNSampled &
  \multicolumn{1}{c}{0.225 $\pm$ 0.018} &
  \multicolumn{1}{c}{0.245 $\pm$ 0.015} &
  0.065 $\pm$ 0.007&
  0.101 $\pm$ 0.001 \\ 
 &
  \RNNConstant &
  \multicolumn{1}{c}{0.288 $\pm$ 0.023} &
  \multicolumn{1}{c}{0.279 $\pm$ 0.024} &
  0.078 $\pm$ 0.004&
  0.087 $\pm$ 0.009 \\ 
 &
  RBU &
  \multicolumn{1}{c}{\textbf{0.098 $\pm$ 0.002}} &
  \multicolumn{1}{c}{\textbf{0.084 $\pm$ 0.002}} &
  \textbf{0.038 $\pm$ 0.001}&
  \textbf{0.034 $\pm$ 0.001} \\ \midrule
  
\multirow{4}{*}{Vegas (Test)} &
    iBoxNet &
  \multicolumn{1}{c}{0.098 $\pm$ 0.000} &
  \multicolumn{1}{c}{0.184 $\pm$ 0.000} &
   0.082 $\pm$ 0.000&
  0.211 $\pm$ 0.000 \\ 
 &
  \RNNSampled &
  \multicolumn{1}{c}{0.254 $\pm$ 0.029} &
  \multicolumn{1}{c}{0.153 $\pm$ 0.038} &
  0.234 $\pm$ 0.012&
  0.125 $\pm$ 0.005 \\ 
 &
  \RNNConstant &
  \multicolumn{1}{c}{0.265 $\pm$ 0.020} &
  \multicolumn{1}{c}{0.145 $\pm$ 0.030} &
  0.270 $\pm$ 0.019&
  0.135 $\pm$ 0.006 \\ 
 &
  RBU &
  \multicolumn{1}{c}{\textbf{0.091 $\pm$ 0.029}} &
  \multicolumn{1}{c}{\textbf{0.089 $\pm$ 0.025}} &
  \textbf{0.036 $\pm$ 0.005}&
  \textbf{0.043 $\pm$ 0.010} \\ \midrule
\end{tabular}
\caption{Wasserstein distances (WD) for \pantheon~data  for different models
and protocols. The lower the better.}
\label{tab:WD_pantheon}
\end{table*}

Next, we quantify how well \buffer~preserves fine-grained temporal patterns. We divide the traces (sending rates, delays) into small chunks (of 15 packets) and compute the maximum mean discrepancy (MMD) between the simulated and the GT chunks, using RBF kernel. We show the MMD (the lower the better) of chunks over time in Figure \ref{fig:rbu-cubic-comparison} for the Vegas (test) protocol. \buffer~has much lower MMD in general, and especially relative to the baseline LSTM models. This suggests that~\buffer~captures local temporal patterns over very long traces. Also, consistent with Table~\ref{tab:WD}, \RNNConstrained~performs better than the baselines, and~\iBoxNet~is competitive. Details on MMD computation and results for different chunk lengths, protocols, and more baselines are in the longer version \cite{ibox_ml}.\\

\textbf{GAN techniques}: As we mentioned in Section~\ref{sec:setup}, GAN techniques~\cite{TimeGAN,jarrett2021time} are unable to scale to sequences of lengths even in the order of hundreds, and for modest model sizes. Setting the output sequence length of the generator to a manageable size, and adversarially training it with aggregate traces, is also not meaningful in our setting --- the output samples from the generator cannot be used to drive simulation as the sender requires continuous packet-level feedback.

\subsection{Simulating real-world network phenomena}
\label{sec:reordering_exp}
We now demonstrate~\buffer's ability to simulate real-world network behaviors using packet re-ordering phenomenon, i.e., packets sent by $S$ arriving out of order at receiver $R$, observed in the~\pantheon traces. This is an important behavior from the application's perspective as reordered packets could be treated as lost if they don't arrive before a certain timeout (depending on the protocol). The metric of interest is the fraction of packets re-ordered in the calls. 

In Figure~\ref{fig:reordering-plots-cubic-vegas}, the metric CDFs for the \buffer~model (with 2 bottleneck links) traces align significantly better with GT, compared to \RNNConstant, for both train and test protocols;~\iBoxNet~is not even shown here since its rigid single FIFO queue model precludes the recreation of reordering. We also show a baseline where we fix the length of the second queue $\tau^{(2)}$ to be twice the first queue $\tau^{(1)}$; this tends to reorder packets flowing through the second (longer) queue; while it performs reasonably well on the train protocol, the match is relatively poor on the test protocol, which underscores the effectiveness of our technique, and the joint learning of the~\buffer~parameters.

\paragraph{Limitations:} To capture and recreate real-world network behaviors such as reordering, we would need domain-specific insights on the new behaviors of interest. It is unlikely that the full expressive power of~\buffer~can be exploited otherwise.

\section{Conclusions}

We formulate a novel ML problem at the intersection of sequential decision making, dynamical systems, and time-series generative modeling. We present the~\buffer~construct that combines domain knowledge with the expressive power of neural models, yielding significantly better match for application-level metrics for network simulation than existing neural techniques and pure domain-knowledge based techniques. We also demonstrate that ~\buffer~is flexible enough to model real-world network phenomena like packet reordering accurately, which is currently not possible using domain-knowledge based techniques like iBoxNet.

\bibliography{aaai23}

\newpage

\appendix
\section{Appendix: Additional Details}
\label{app:details}
\subsection{~\buffer~learning}
\label{app:algo}
The procedure for learning the~\buffer~model, for the single-bottleneck link case, is given in Algorithm~\ref{algo:learnRBU}. We initialize $\ThetaWin$ weights randomly, and use the heuristic initialization strategy discussed in Section~\ref{sec:exp} for $\ThetaRBU$. The hyper-parameters $\lambda$, $\gamma$ are specified/tuned as discussed in Appendix~\ref{app:hyperparams}. For the~\ns dataset, we set $N_w = 600$, corresponding to chunking 60-second calls into 100-millisecond windows. For the~\pantheon dataset, $N_w = 300$, corresponding to chunking 30-second calls into 100-millisecond windows.

\begin{algorithm*}[t]
	\begin{algorithmic}[0]
        
		\STATE \textbf{Input:} traces $\gT = \{(\rvx^{(i)}_t, y^{(i)}_t)_{t=1}^{\tracelen_i}\}_{i=1}^\numtraces$, $\gamma$, $\lambda$, initial $\ThetaWin^{(0)}, \ThetaRBU^{(0)}$ 
		\STATE \textbf{Output: } Model parameters $\ThetaWin, \ThetaRBU$
		\item[]
		\STATE \textbf{Init:} $\ThetaWin = \ThetaWin^{(0)}$, $\ThetaRBU = \ThetaRBU^{(0)} := (g^{(0)}, W_h^{(0)}, U_h^{(0)}, \rvw_c^{(0)})$\\
		\FOR{epochs $ 1 \leq e \leq e_{\text{max}}$ and mini-batches $\gB \in \gT$}
		\STATE \algorithmiccomment{Update trace-level \emph{$\buffer$} parameters}
		\STATE $\big(\tau^{(i)}, \propd^{(i)}, \draind^{(i)}\big) = g(\rvx^{(i)})$, for traces $i \in \gB$
		\FOR{window $w \leq 1 \leq \numwindows$ and trace $i \in \gB$}
        \STATE \algorithmiccomment{Perform forward pass on the window model}
		\STATE $(\rvh^{(i)}_w, c^{(i)}_w) = \text{LSTM}(\rvh^{(i)}_{w-1}; \rvx^{(i)}, \ThetaWin)$
        \STATE \algorithmiccomment{Compute window-level loss}
		\STATE 
		$\ell^{(i,w)}_{\text{window}} \leftarrow \ell_{\text{CE}}(c_w^{(i)}, \tilde{\rvc}^{(i)}_w)$, where $\tilde{\rvc}^{(i)}_w$ is as used in Equation~\eqref{eqn:windowloss}
		\STATE \algorithmiccomment{Perform forward pass of \emph{\buffer} model}
		\STATE $\rvh^{(i)}_t = \sigma\big(\langle W_h, \big[\rvh^{(i)}_w \ \rvx^{(i)}_t\big] \rangle + \langle U_h, \rvh_{t-1}\rangle\big)$, for $t \in \text{window } w$
		\STATE Update $\queued_t$ as in~\eqref{eqn:qt} and $c_t$ as in~(\ref{eqn:CTupdated}), using $\big(\tau^{(i)}, \propd^{(i)}, \draind^{(i)}, c_w^{(i)}\big)$, for $t \in \text{window } w$
		\STATE Predict $\hat{y}^{(i)}_t$ and $p^{(i)}_t$ as in~(\ref{eqn:RBU}), for $t \in \text{window } w$.
		\STATE \algorithmiccomment{Compute packet-level losses}
		\STATE $\ell_{\text{pkt}}^{(i,w,t)} \leftarrow  \ell_{\text{pkt}}\big((\hat{y}^{(i)}_t, p_t^{(i)}), y_t^{(i)}\big)$ as in Equation~\eqref{eqn:packetloss}
		\ENDFOR
    \STATE \algorithmiccomment{Perform back prop}
    \STATE $\nabla_{\{\ThetaWin,\ThetaRBU\}}\ell_{\text{pkt}} := \displaystyle\frac{1}{|\gB|}\frac{1}{\numwindows} \sum_{i \in \gB}\sum_{w \in \numwindows} \frac{1}{|\{t \in w\}|}\sum_{t \in w}  \nabla  \ell_{\text{pkt}}(\ \cdot\ ; \ThetaRBU)_{|\ell_{\text{pkt}}^{(i,w,t)}}$
    \STATE $\nabla_{\{\ThetaWin,\ThetaRBU\}}\ell_{\text{window}} := \displaystyle\frac{1}{|\gB|}\frac{1}{\numwindows} \sum_{i \in \gB}\sum_{w \in \numwindows} \nabla \ell_{\text{window}}(\ \cdot\ ; \ThetaWin)_{|\ell_{\text{window}}^{(i,w)}}$    
    \STATE \algorithmiccomment{Update model parameters}
	\STATE $\ThetaWin \leftarrow \ThetaWin - \eta \nabla_{\ThetaWin}\ell_{\text{window}} - \lambda \eta \nabla_{\ThetaWin}\ell_{\text{pkt}}$   
	\STATE $\ThetaRBU \leftarrow \ThetaRBU - \eta \nabla_{\ThetaRBU}\ell_{\text{window}} - \lambda \eta \nabla_{\ThetaRBU}\ell_{\text{pkt}}$   	
	\ENDFOR 
	\end{algorithmic}
	\caption{Learning the proposed \bufferdesc{} model}
	\label{algo:learnRBU}
\end{algorithm*}



\subsection{Implementation details \& hyperparameter selection}
\label{app:hyperparams}
\paragraph{Data preparation for different models.} Each static feature in $\rvx$ is normalized between 0 and 1 on the entire dataset, and we use this normalized feature during both training and at inference, for all the models. We normalize the input $\rvx_t$ (e.g., window-level aggregate sending rates,  delay values in case of the baselines, or $\tilde{\rvc}_w$ in case of the window-model for~\buffer) to the LSTM models \RNNConstant,~\RNNSampled, and~the window-model of~\buffer, between 0 and 1 using the max values of the corresponding trace, for training. At inference, we denormalize the output values of the LSTM (which is a distribution over 100 binned delays) as follows. A value is uniformed sampled from the distribution of max values for the corresponding feature across all training traces, and is used to scale the sampled value.
 
 We discretize the output of \RNNConstant~and~\RNNSampled, and the window-model of \buffer~into 100 equisized bins. The LSTM outputs a multinomial distribution over 100 bins where each bin corresponds to a range of percentage values. At inference time, we recover a continuous range by first sampling the bin according to the multinomial distribution output by the model, and then sampling uniformly from each bin. We observe that this discretization step improves fidelity of the generated traces as compared to using a parameterized continuous distribution such as Gaussian which is far less representative.
 
 We similarly discretize inputs to the GPT decoder~\cite{radford2018improving} used in the \textbf{Transformer} baseline. It receives a sequence of discretized delays ($y_{t}$) and inter-packet gaps ($s_{t}$), and predicts the (discretized) delay experienced in the next time step. During inference (simulation), we provide an initial packet delay and the resulting inter-packet gap as context to the decoder. The initial (discretized) delay is sampled from a multinomial distribution over 100 bins, where each bin corresponds to some initial delay of the training data, normalized by the maximum delay values of their corresponding traces. Of the traces corresponding to a given bin, the maximum delay value is sampled and is used to scale the delay values returned by the decoder. 
 
\paragraph{Hyperparameter tuning.} For the baselines \RNNConstant~and~\RNNSampled, we use a standard LSTM~\citep{sutskever2011generating, DBLP:journals/corr/Graves13} implemented in PyTorch with $2$ hidden layers, hidden size of $256$, and a fully connected layer with 100 output dimensions corresponding to the discretized total delay and a softmax layer. The architecture was tuned to maximize the mean delay and throughput distribution match with the ground-truth on the training (Cubic) protocol. 

For the window-level model of~\buffer, we use the same LSTM architecture as above, where the output corresponds to discretized $c_w$ (used in~\ref{eqn:CTupdated}) and $q_w$. We set hidden dimension, for $\bm{h}_t$ in \eqref{eqn:CT}, as 1 which works without much tuning across datasets. $W_{h}$ is a fully connected layer with an input dimension of $5$ and output dimension of $1$ and $U_{h}$ is just a scalar. We set $\gamma$ in \eqref{eqn:CTupdated} to $0.1$ in all our experiments. We set the learning rate $\eta$ to $0.001$ for the window-level model and to $0.01$ for the packet-level model, and $\lambda = 1$ in Algorithm~\ref{algo:learnRBU}.

For the \textbf{Transformer} baseline, we use a standard GPT model (decoder) implemented in PyTorch with $2$ layers and a context window of size $128$. We utilize $4$ attention heads and an embedding dimension of $256$. We used the AdamW \footnote{Refer https://openreview.net/pdf?id=Bkg6RiCqY7} optimizer to train the model, with an initial learning rate of $0.0006$.

\paragraph{Efficiency of \buffer~(Scaling to high-bandwidth networks):} Our method utilizes a window-level model (\RNNConstant) with 1.9M parameters and a packet-level~\buffer~model with just 32 parameters. Our baselines  and~\RNNSampled~and~\RNNConstrained~are identical to~\RNNConstant~and have 1.9M parameters as well.

The LSTM used in the~\RNNSampled~and~\RNNConstrained~baselines has an inference time of around 2ms per packet. In general, it can be very challenging to use these models to emulate networks of high bottleneck bandwidths ($\sim$15Mbps). However,~\buffer~can support the dynamics of such high bandwidth networks, as described below.

As noted above, we use the same LSTM architecture for the window level model as the baselines. However, in our approach, the LSTM can be unrolled ahead of time as it works on a coarse window level granularity using only the static features (as mentioned in \textbf{``RBU Inference"}, Section 4). Thus, the cost of window model inference can be completely amortized. The packet level  model is quite compact with only 32 parameters; thanks to the way we designed the update equations using domain knowledge. Its forward pass can be accomplished typically within 0.5ms.

This implies that, in theory, we should be able to handle networks with a bottleneck bandwidth of around 24Mbps (assuming a typical packet of 1.5KB). We are certain that better engineering and the usage of a GPU for inference can only improve latencies, especially when there are packet bursts.

\paragraph{Multi-path~\buffer~model details.} For experiments on the~\pantheon~data, we use a 2-path~\buffer~model. The size of the second queue is expressed as a scaling of the first queue. To estimate this scaling factor, we use a similar $g$ model as in~\eqref{eqn:staticg} with static features as input. We use a window-level $q_w$ obtained as discussed in Section~\ref{sec:bufferunit}. At training time, the $q_{w}$ obtained as the $\argmax$ from the window-model output (just like in the case of $c_w$ for the window-level cross-traffic model), is used as the probability of choosing the second queue and is used to calculate the expected delay for a given packet for all the packets in the window $w$. However, at inference, we replace the $\argmax$ with actual sampling, obtain a $q_{w}$ for the current window, and then for every packet in the window, we choose one of the two queues, based on sampling from $\text{Bernoulli}(q_{w})$, and compute the delay based on the corresponding queue dynamics.

\subsection{Additional Results}
\label{app:results}

\textbf{Metrics: }We evaluate the simulation methods on three types of distributional metrics:
\begin{enumerate}
    \item \textbf{Wasserstein distance (WD)}: Wasserstein distance is a standard measure of discrepancy between ground truth and model generated distributions, but calculating the Wasserstein distance between high dimensional distributions is intractable. However, network applications (such as video call conferencing) typically are interested in certain marginal and joint distributions of trace-level (aggregate) quantities. So, we look at distributions of trace-level throughput (mean receiving rate) and delays (P95 and mean). We compute a 2D Wasserstein-2 distance (also called earth mover's distance) for the joint distribution between the mean throughput and 95th percentile delay in Table \ref{tab:appendixresults} and \ref{tab:WD_pantheon}, and joint distribution between the mean throughput and mean delay in Table \ref{tab:WD} and \ref{tab:WD_pantheon}. To be able to compute the distance, we normalize the trace-level aggregate throughput and delay quantities by a factor of $\frac{1}{\text{throughput}_{\text{max}} - \text{throughput}_{\text{min}}}$ and $\frac{1}{\text{delay}_{\text{max}} - \text{delay}_{\text{min}}}$ respectively, where $\text{throughput}_{\text{max}}$ ($\text{throughput}_{\text{min}}$) denotes the maximum (minimum) mean throughput achieved across all the traces in all the distributions and $\text{delay}_{\text{max}}$ ($\text{delay}_{\text{min}})$ denotes maximum and minimum delay observed across all the traces in all the distributions.
    We also compute 1D Wasserstein scores for individual trace-level aggregate quantities like mean delay and 95th percentile delay which can be interpreted as the area between the cdfs of the two distributions being compared. We use \texttt{scipy.stats.wasserstein\_distance} from the Scipy library for the 1D Wasserstein distance and $\texttt{wasserstein.emd()}$ from the Wasserstein python package \footnote{\url{https://pkomiske.github.io/Wasserstein/docs/emd/}} for the 2D Wasserstein distance.

    \item \textbf{ Maximum mean discrepancy (MMD)}: 
    Maximum mean discrepancy or MMD quantifies the distance between distributions in terms of a kernel function. In this work, we use the  RBF kernel of the form $k(\bm{x}, \bm{x}) = \exp(-\zeta \| \bm{x} - \bm{y} \|^2)$ for appropriately small $\zeta$ (in our evaluation, we set $\zeta = 1$ for the~\ns~data and $\zeta = 0.1$ for the \pantheon~data). For this kernel, MMD between distributions $P$ and $Q$ can be defined as: 
    $ MMD^2(P, Q) = E_{\bm{x}, \bm{x}' \sim P}[k(\bm{x}, \bm{x}')] + E_{\bm{x}, \bm{x}' \sim Q}[k(\bm{x}, \bm{x}')] - 2E_{\bm{x} \sim P, \bm{x} \sim Q}[k(\bm{x}, \bm{x})] $ \\
    We use this metric to compute the closeness of match between the model generated traces and real traces at a fine-grained, temporal level. We first normalize each dimension of each trace between 0 and 1 (by min-max normalization) and truncate or pad traces to meet the length of the ground truth traces. We first divide a trace into chunks of 50 packets each starting at packet number 0, 100, 200, and so on. We then further divide each of these chunks into mini-chunks of length $n = 15$ with 3 dimensions (packet loss, delay $y_t$, and inter-packet spacing $s_t$) and convert each of these chunks into a vector of  length $3n$. \footnote{we also tried a few other small values for $n$, and the results are consistent}  Thus, each chunk can be transformed into a set of vectors of length $3n$ associated with a starting packet number $i$, and these sets are aggregated across traces of the same static configurations keeping the starting packet number of the chunk the same. We then compute the MMD score between the two sets of vectors (corresponding to the chunks starting from packet number $i$ and derived from the real and the generated traces) for every static configuration and then average across the static configurations to get a single score for each chunk number $i$. For \pantheon data where we do not have any information on the underlying network configurations, we simply compute an aggregate score over all traces. We plot this averaged MMD score as a function of chunk number $i$ which gives a quantification of how well the different methods preserve the fine-grained structure of traces over time. 
    
    \item \textbf{Discriminative score}: Following the implementation in the evaluation section of \citet{TimeGAN}, we train a 2-layer GRU with tanh activation and 2 hidden units to classify the model generated traces from ground-truth traces using a subset of traces for training the GRU, and evaluate the performance of the trained GRU on held-out datasets. If the classifier achieves an accuracy of 1, it means that the datasets are perfectly distinguishable by the discriminator; if it achieves an accuracy of 0.5, the datasets are indistinguishable from each other. We subtract 0.5 from the test accuracies to obtain a ``discriminative score'' (the lower the better). We repeat this for the simulated traces obtained using different models, and compute the discriminative scores in each case. The discriminator GRU architecture is fixed across the compared simulation methods, which makes this evaluation fair.
\end{enumerate}

Table \ref{tab:appendixresults} shows discriminative scores, 2D-wasserstein distances for joint throughput, P95 delay distribution, and  1D-wasserstein distances for mean delay distribution, for the compared methods, on~\ns~data. The results are consistent with our observations from Table~\ref{tab:WD} in the main paper. Table~\ref{tab:WD_pantheon} shows the Wasserstein distances on the~\pantheon~dataset. Here, we find that our~\buffer~model consistently outperforms all the compared methods.  Note that we do not try \RNNConstrained~on the~\pantheon~dataset as these real-world traces naturally have reordered packets.

\begin{figure*}[ht!]
    \centering
    \includegraphics[width=\textwidth]{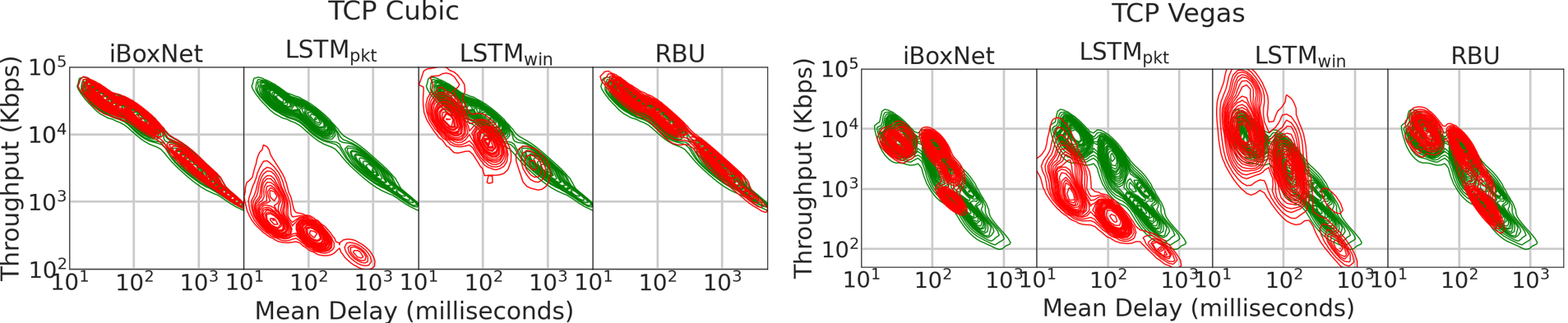}
    \caption{Contour plots for the joint distribution of mean delay and throughput in the ground-truth traces (green) and the traces produced by the~\buffer~model (red) trained with TCP Cubic data and tested on TCP Cubic (left) and Vegas (right).}
    \label{fig:all-protocols-comparison}
\end{figure*}

\begin{figure*}
    \centering
    \includegraphics[width=1.02\textwidth]{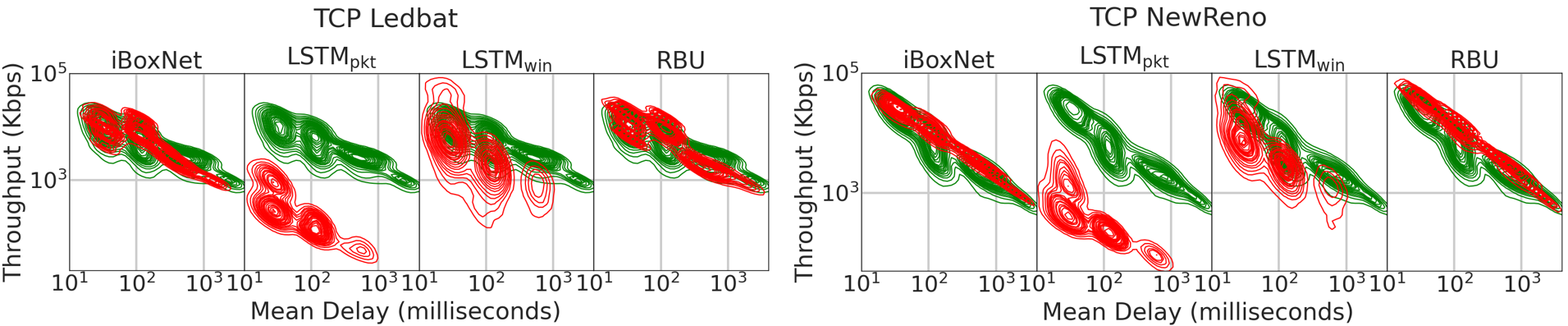}
    \caption{Contour plots for the joint distribution of mean delay and throughput in the ground-truth traces (green) and the traces produced by the~\buffer~model (red) trained with TCP Cubic data and tested on TCP Ledbat (left) and NewReno (right).}
    \label{fig:all-protocols-comparison-appendix}
\end{figure*}

Figures \ref{fig:all-protocols-comparison} and~\ref{fig:all-protocols-comparison-appendix} show the joint distribution of mean delay and throughput in the ground-truth traces and the simulated traces obtained by the compared methods. We observe that the simulated traces obtained by \buffer~match the ground-truth traces quite well in terms of the joint delay, throughput distribution, in both the train and the test protocols. Furthermore, we notice how the~\buffer~model captures the three clusters in the ground truth, pertaining to three types of network configurations (the 3 scenarios in Table~\ref{tab:settings}), low delay-high throughput, moderate delay-moderate throughput and high delay-low throughput. 

\begin{figure*}

\centering
     \includegraphics[width=0.45\textwidth]{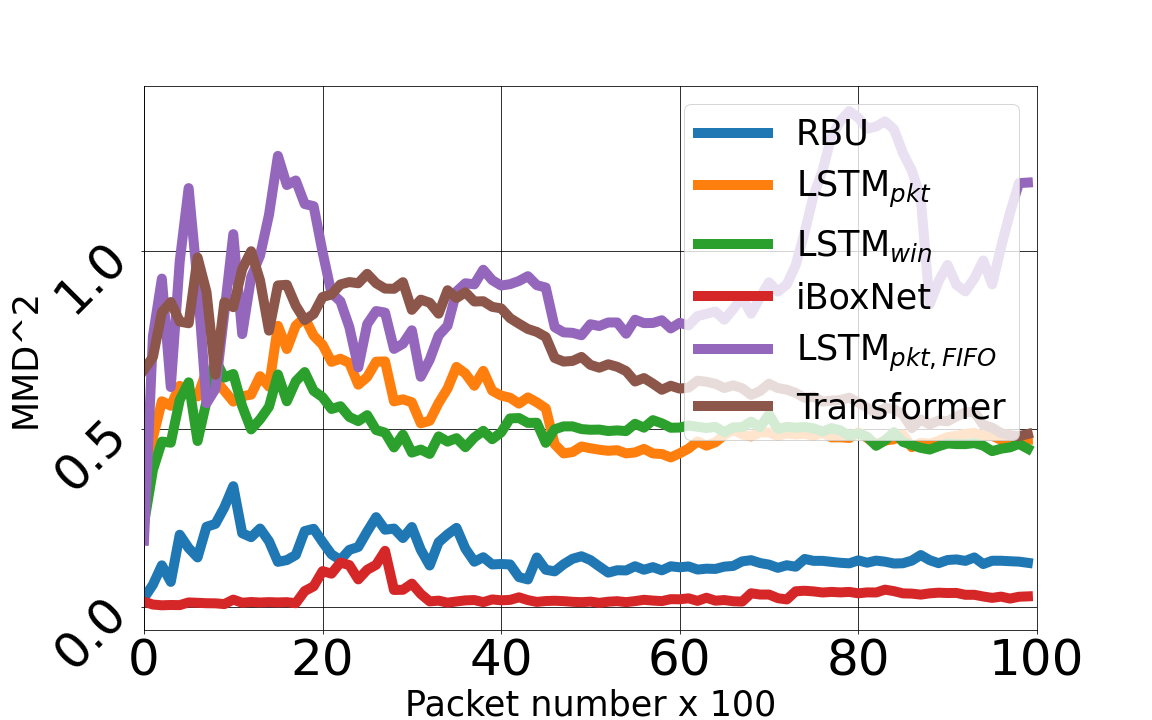}
    \includegraphics[width=0.51\textwidth]{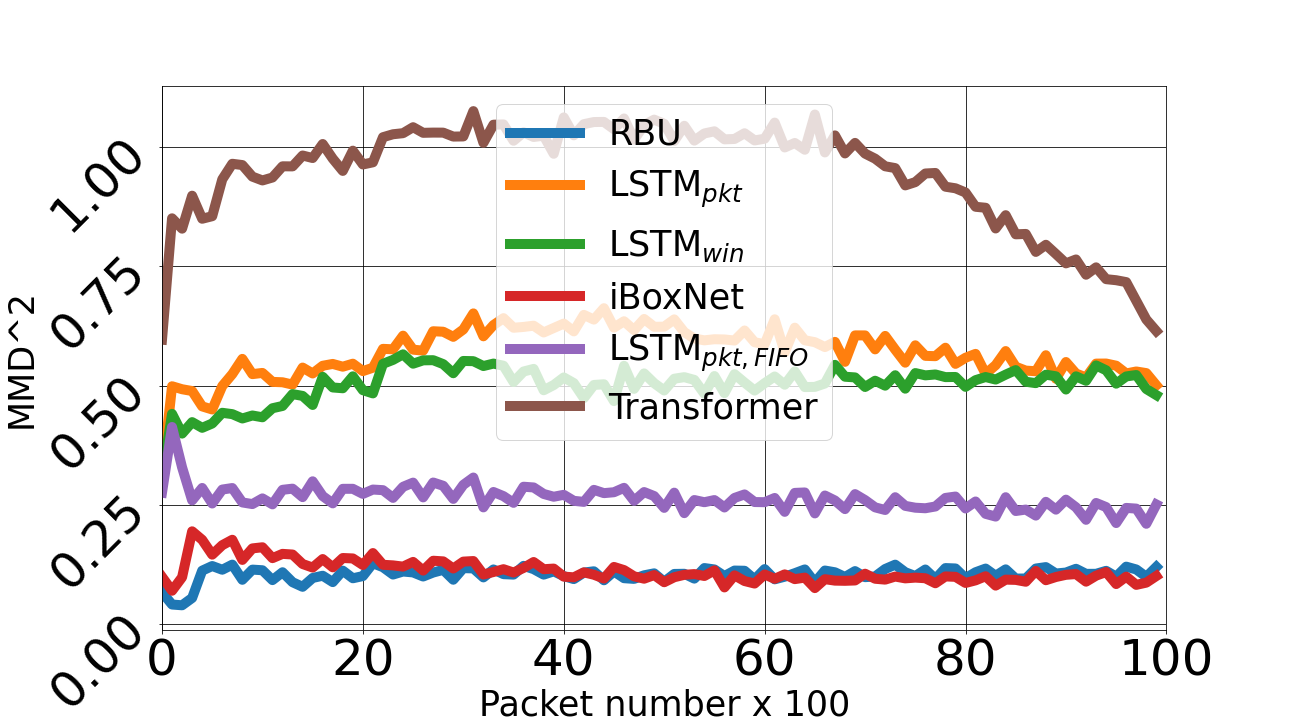}

    \caption{$MMD^2$ (with $\zeta = 1$) vs starting packet number (of chunks) on the \ns dataset for (left) the training protocol Cubic and (right) the test protocol Vegas.}
    \label{fig:mmd_synthetic}
\end{figure*}

\begin{figure*}

\centering
    \includegraphics[width=0.45\textwidth]{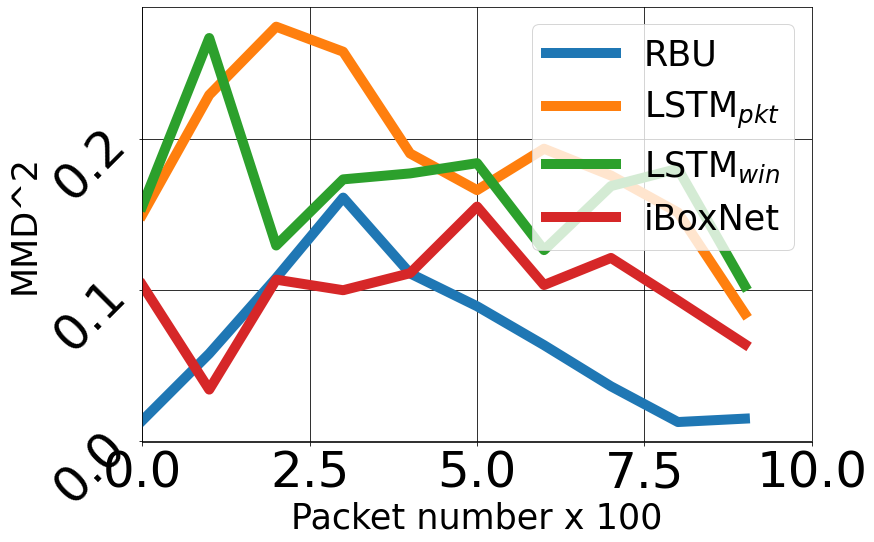}
    \includegraphics[width=0.45\textwidth]{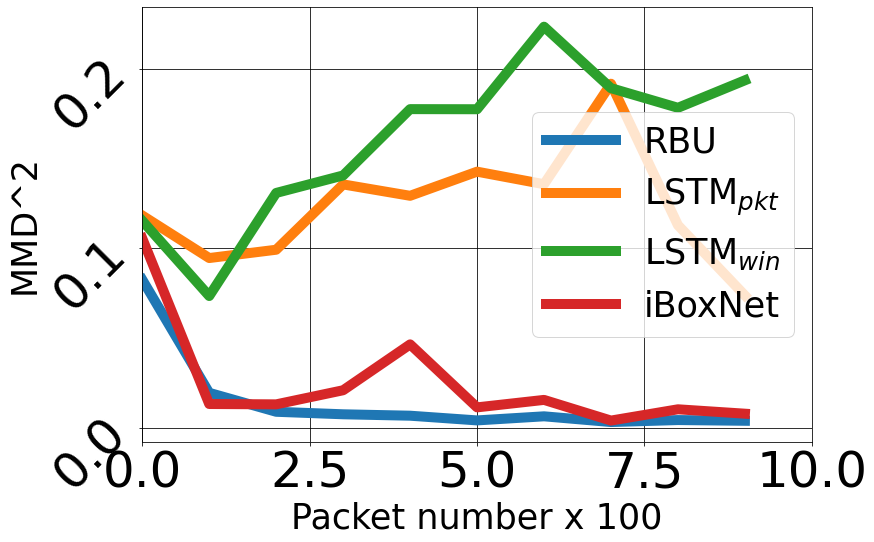}

    \caption{$MMD^2$ (with $\zeta= 0.1$) vs starting packet number (of chunks) on the \pantheon dataset for (left) the training protocol Cubic and (right) the test protocol Vegas.}
    \label{fig:mmd_pantheon}
\end{figure*}

Figures~\ref{fig:mmd_synthetic} and \ref{fig:mmd_pantheon} show the chunk-wise MMD-based metric computed as described above vs packet number (of the beginning of a chunk) for the compared methods, on (train) Cubic and (test) Vegas protocols, for~\ns~and~\pantheon~datasets respectively. The results are consistent with our observations from Figure~\ref{fig:rbu-cubic-comparison} in the main paper. Note that, as stated earlier, we do not try \RNNConstrained~on the~\pantheon~dataset as these real-world traces naturally have reordered packets.

\textbf{Comparison with Transformers.} Table~\ref{tab:appendixresults_transformer} shows the Wasserstein distances obtained using the Transformer model~\cite{radford2018improving} trained on Cubic protocol, and tested on all the protocols, on the~\ns~data. Comparing the values with the corresponding numbers in Tables~\ref{tab:WD} and~\ref{tab:appendixresults} (the lower the better), we find that~\buffer~consistently and significantly outperforms the Transformer model in almost all cases (except the two values shown in \textbf{bold}), even though the Transformer model performs better than the LSTM baselines in many cases.

\begin{table*}[ht]
\centering
\small \begin{tabular}{l l r r r}
\toprule
\multirow{2}{*}{Protocol} &
  \multirow{2}{*}{Model (vs GT)} &
  \multirow{2}{*}{Disc Score} &
  \multicolumn{2}{c}{Wasserstein Distances} \\ 
 &
   &
   &
2D (Tput, P95 Delay) &
1D Mean Delay \\ \toprule
\multirow{6}{*}{Cubic (Train)} &
  GT &
  0.006 $\pm$ 0.005 &
  \multicolumn{1}{c}{-} &
  \multicolumn{1}{c}{-} \\ 
 &
  iBoxNet &
  \textbf{0.076 $\pm$ 0.029} &
  \multicolumn{1}{c}{\textbf{0.013 $\pm$ 0.000}} &
  \textbf{0.002 $\pm$ 0.000} \\ 
 &
  \RNNSampled &
  0.499 $\pm$ 0.001 &
  \multicolumn{1}{c}{0.283 $\pm$ 0.011} &
  0.144 $\pm$ 0.001 \\
 &
  \RNNConstant &
  0.463 $\pm$ 0.010 &
  \multicolumn{1}{c}{0.177 $\pm$ 0.002} &
  0.138 $\pm$ 0.001 \\ 
 &
  \RNNConstrained &
  0.442 $\pm$ 0.046 &
  \multicolumn{1}{c}{0.224 $\pm$ 0.009} &
  0.108 $\pm$ 0.000 \\ 
 &
  RBU &
  0.236 $\pm$ 0.024 &
  \multicolumn{1}{c}{0.032 $\pm$ 0.004} &
  \textbf{0.007 $\pm$ 0.000} \\ \midrule
\multirow{6}{*}{Vegas (Test)} &
  GT &
  0.010 $\pm$ 0.008 &
  \multicolumn{1}{c}{-} &
  \multicolumn{1}{c}{-} \\ 
 &
  iBoxNet &
  0.485 $\pm$ 0.001 &
  \multicolumn{1}{c}{0.094 $\pm$ 0.000} &
  0.044 $\pm$ 0.000 \\ 
 &
  \RNNSampled &
  0.480 $\pm$ 0.001 &
  \multicolumn{1}{c}{0.124 $\pm$ 0.001} &
  0.075 $\pm$ 0.001 \\
 &
  \RNNConstant &
  0.470 $\pm$ 0.002 &
  \multicolumn{1}{c}{0.135 $\pm$ 0.002} &
  0.074 $\pm$ 0.002 \\
 &
  \RNNConstrained &
  0.497 $\pm$ 0.000 &
  \multicolumn{1}{c}{0.104 $\pm$ 0.000} &
  0.046 $\pm$ 0.000 \\
 &
  RBU &
  0.481 $\pm$ 0.002 &
  \multicolumn{1}{c}{\textbf{0.073 $\pm$ 0.002}} &
  \textbf{0.022 $\pm$ 0.009} \\ \midrule
\multirow{6}{*}{LEDBAT (Test)} &
  GT &
  0.009 $\pm$ 0.005 &
  \multicolumn{1}{c}{-} &
  \multicolumn{1}{c}{-} \\ 
 &
  iBoxNet &
  0.266 $\pm$ 0.004 &
  \multicolumn{1}{c}{\textbf{0.023 $\pm$ 0.000}} &
  0.039 $\pm$ 0.000 \\ 
 &
  \RNNSampled &
  0.499 $\pm$ 0.001 &
  \multicolumn{1}{c}{0.198 $\pm$ 0.001} &
  0.112 $\pm$ 0.001 \\
 &
  \RNNConstant &
  0.476 $\pm$ 0.001 &
  \multicolumn{1}{c}{0.161 $\pm$ 0.001} &
  0.106 $\pm$ 0.000 \\ 
 &
  \RNNConstrained &
  0.433 $\pm$ 0.037 &
  \multicolumn{1}{c}{0.145 $\pm$ 0.001} &
  0.080 $\pm$ 0.000 \\ 
 &
  RBU &
  \textbf{0.216 $\pm$ 0.023} &
  \multicolumn{1}{c}{0.041 $\pm$ 0.001} &
  \textbf{0.016 $\pm$ 0.001} \\ \midrule
\multirow{6}{*}{NewReno (Test)} &
  GT &
  0.008 $\pm$ 0.002 &
  \multicolumn{1}{c}{-} &
  \multicolumn{1}{c}{-} \\ 
 &
  iBoxNet &
  \textbf{0.235 $\pm$ 0.014} &
  \multicolumn{1}{c}{\textbf{0.037 $\pm$ 0.000}} &
  \textbf{0.023 $\pm$ 0.000} \\ 
 &
  \RNNSampled &
  0.499 $\pm$ 0.001 &
  \multicolumn{1}{c}{0.235 $\pm$ 0.009} &
  0.125 $\pm$ 0.001 \\ 
 &
  \RNNConstant &
  0.480 $\pm$ 0.010 &
  \multicolumn{1}{c}{0.165 $\pm$ 0.006} &
  0.115 $\pm$ 0.007 \\ 
 &
  \RNNConstrained &
  0.348 $\pm$ 0.008 &
  \multicolumn{1}{c}{0.189 $\pm$ 0.007} &
  0.095 $\pm$ 0.000 \\ 
 &
  RBU &
  0.266 $\pm$ 0.017 &
  \multicolumn{1}{c}{0.097 $\pm$ 0.020} &
  \textbf{0.023 $\pm$ 0.011} \\ \bottomrule
\end{tabular}
\caption{Additional results for Table~\ref{tab:WD}: (mean $\pm$ std. dev) Discriminative scores, Wasserstein distances (WD), the lower the better, for different models and protocols, on the~\ns~data.}
\label{tab:appendixresults}
\end{table*}

\begin{table*}[ht]
\centering
\small \begin{tabular}{l r r r r}
\toprule
\multirow{2}{*}{Protocol} &
  \multicolumn{4}{c}{Wasserstein Distances} \\ 
   &
   2D (Tput, Mean Delay) &
2D (Tput, P95 Delay) &
1D Mean Delay &

1D P95 Delay \\ \toprule
Cubic (Train) &
  0.224 $\pm$ 0.015 &
  0.207 $\pm$ 0.016 &
  0.048 $\pm$ 0.003&
  0.030 $\pm$ 0.003 \\ 
  
Vegas (Test) &
  0.079 $\pm$ 0.003 &
  \textbf{0.032} $\pm$ \textbf{0.001} &
   0.044 $\pm$ 0.003&
  \textbf{0.007 }$\pm$ \textbf{0.001} \\ 

LEDBAT (Test) &
  0.103 $\pm$ 0.001 &
  0.110 $\pm$ 0.005 &
  0.036 $\pm$ 0.001&
  0.043 $\pm$ 0.005 \\ 

Newreno (Test) &
  0.165 $\pm$ 0.013 &
  0.166 $\pm$ 0.012 &
   0.036 $\pm$ 0.002&
  0.038 $\pm$ 0.002 \\ \bottomrule

\end{tabular}
\caption{Additional results for Table~\ref{tab:WD}: (mean $\pm$ std. dev) Wasserstein distances (WD) for the \textbf{Transformer} model~\cite{radford2018improving}, for different protocols on the~\ns~data. Comparing the values with the corresponding numbers in Tables~\ref{tab:WD} and~\ref{tab:appendixresults} (the lower the better), we find that~\buffer~consistently and significantly outperforms the Transformer model in almost all cases (except the two values shown in \textbf{bold}), even though the Transformer model performs better than the LSTM baselines in many cases.}
\label{tab:appendixresults_transformer}
\end{table*}

\section{Appendix: Dataset Details}

\label{app:datasets}
\subsection{\ns}
\label{app:ns3data}







For generating the synthetic dataset, we randomly sample 14 link configurations from each of the three network scenarios as shown in Table \ref{tab:settings}. For each of the $42$ sampled $(14 \times 3)$ link configurations, we generate 70 cross-traffic patterns keeping the link configuration fixed, yielding a total of $2940$ $(42 \times 70)$ traces. We generate such a set of $2940$ traces for each of the four protocols: TCP Cubic, Vegas, LEDBAT and NewReno, to use as the ground-truth in our evaluation. For the purpose of training, we sample $31$ Cubic traces from each of the $42$ link configurations and use the resulting $1302$ $(42 \times 31)$ traces as the training data. 
\begin{table*}
\centering
\small \begin{tabular}{@{}c c c c}
\toprule
Network Scenario & Bottleneck Bandwidth & Propagation Delay & Buffer Size \\
\toprule
Scenario 1 & 40-50 Mbps & 20-50 ms & 20-50 MTU packets   \\
Scenario 2 & 20-30 Mbps & 90-120 ms & 100-150 MTU packets\\
Scenario 3 & 1-10 Mbps & 150-200 ms & 300-500  MTU packets \\
\bottomrule 
\end{tabular}
\caption{Network configuration settings for a dumbbell topology on $\ns$, for generating synthetic traces used in this paper. MTU refers to Maximum Transfer Unit, i.e., the maximum size of a packet on the network.}
\label{tab:settings}
\end{table*}

\subsection{\pantheon}
\label{app:pantheondata}
Pantheon \cite{yan2018pantheon} is a public dataset consisting of packet level network traces obtained from a number of vantage points across the globe. The traces cover a range of protocols (including TCP Cubic and Vegas, which we focus on here) and the vantage points cover different network types (e.g., Ethernet, cellular, etc.). The results presented in this paper are based on the $47$ traces obtained from the China cellular network. 

\section{Appendix: Example Traces}
\label{app:ns3}
\begin{figure*}[ht!]
    \centering
    \includegraphics[width=1.02\textwidth]{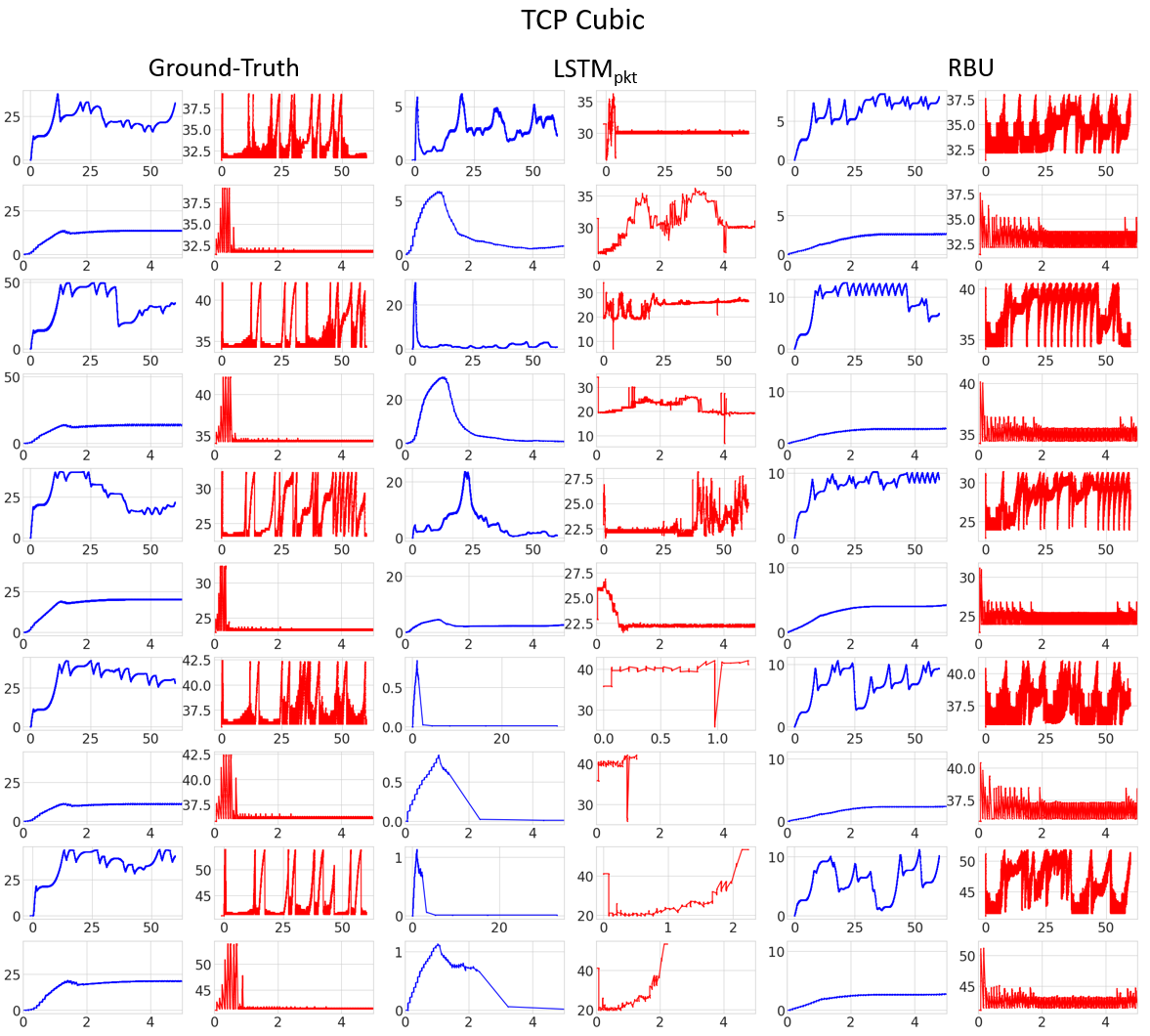}
    \caption{Traces produced by~\RNNSampled~(middle) and the~\buffer~(right) models along with Ground-Truth traces (left) for the test protocol TCP Cubic. The blue curves represent sending rates decided by the test protocol and the red curves represent delays generated by the models. For every pair of rows, the first row shows the entire trace and the second row focuses on the first few seconds. These are randomly picked traces, and there is no one-to-one correspondence between columns.}
    \label{fig:sample-traces-cubic}
\end{figure*}

\begin{figure*}[ht!]
    \centering
    \includegraphics[width=1.02\textwidth]{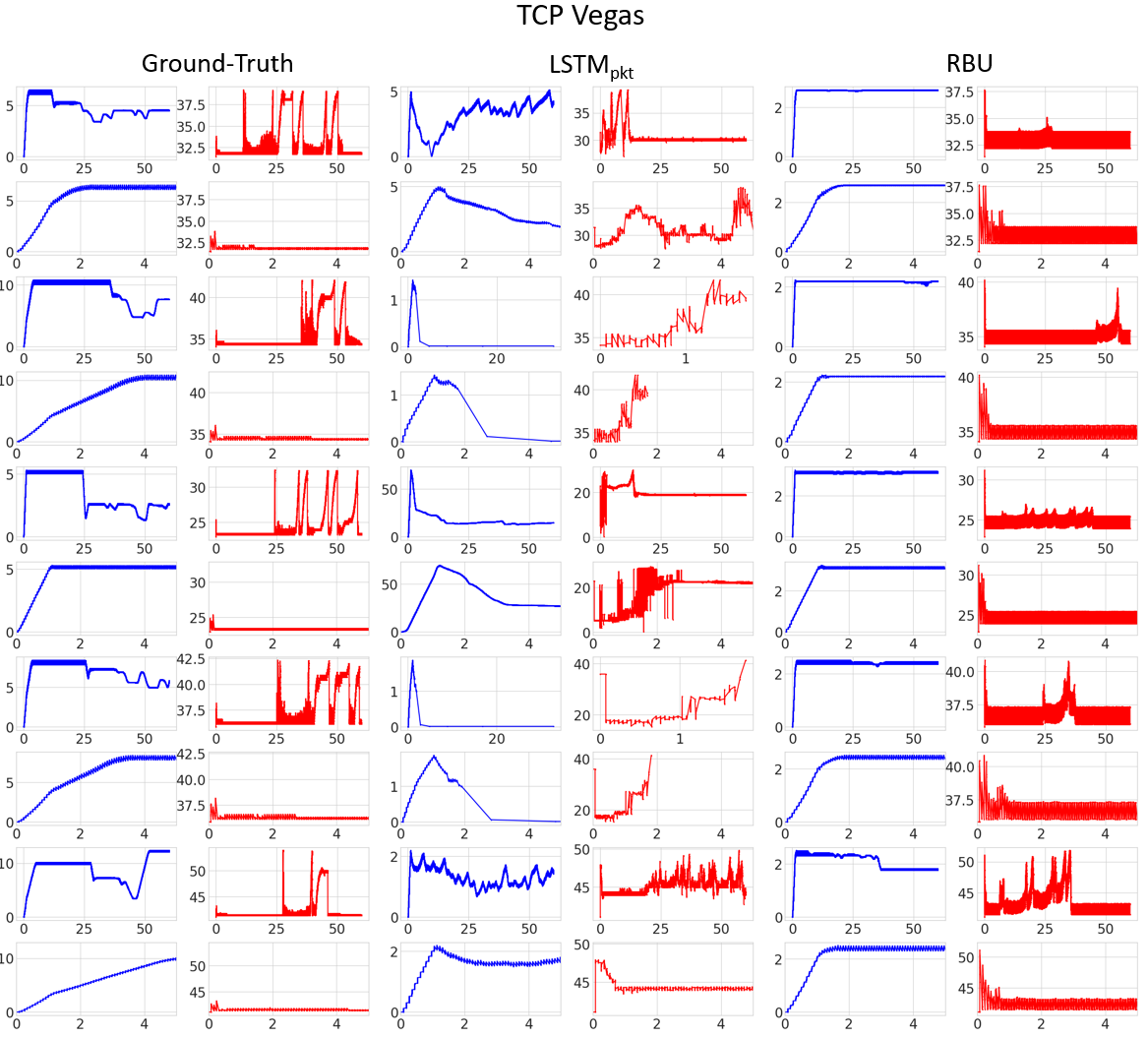}
    \caption{Traces produced by~\RNNSampled~(middle) and the~\buffer~(right) models along with Ground-Truth traces (left) for the test protocol TCP Vegas. The blue curves represent sending rates decided by the test protocol and the red curves represent delays generated by the models. For every pair of rows, the first row shows the entire trace and the second row focuses on the first few seconds. These are randomly picked traces, and there is no one-to-one correspondence between columns.}
    \label{fig:sample-traces-vegas}
\end{figure*}

\begin{figure*}[ht!]
    \centering
    \includegraphics[width=1.02\textwidth]{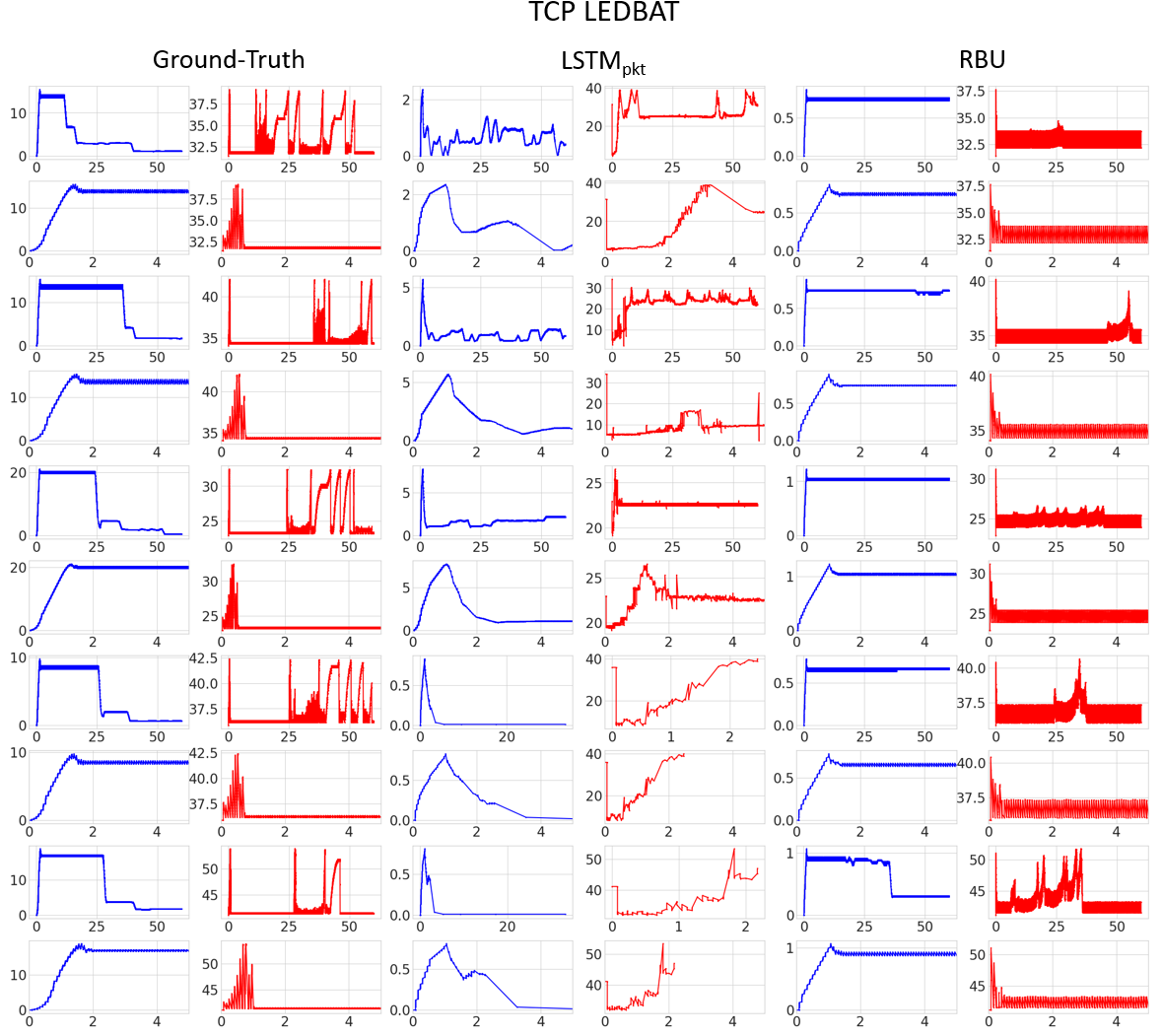}
    \caption{Traces produced by~\RNNSampled~(middle) and the~\buffer~(right) models along with Ground-Truth traces (left) for the test protocol TCP LEDBAT. The blue curves represent sending rates decided by the test protocol and the red curves represent delays generated by the models. For every pair of rows, the first row shows the entire trace and the second row focuses on the first few seconds. These are randomly picked traces, and there is no one-to-one correspondence between columns.}
    \label{fig:sample-traces-ledbat}
\end{figure*}

\begin{figure*}[ht!]
    \centering
    \includegraphics[width=1.02\textwidth]{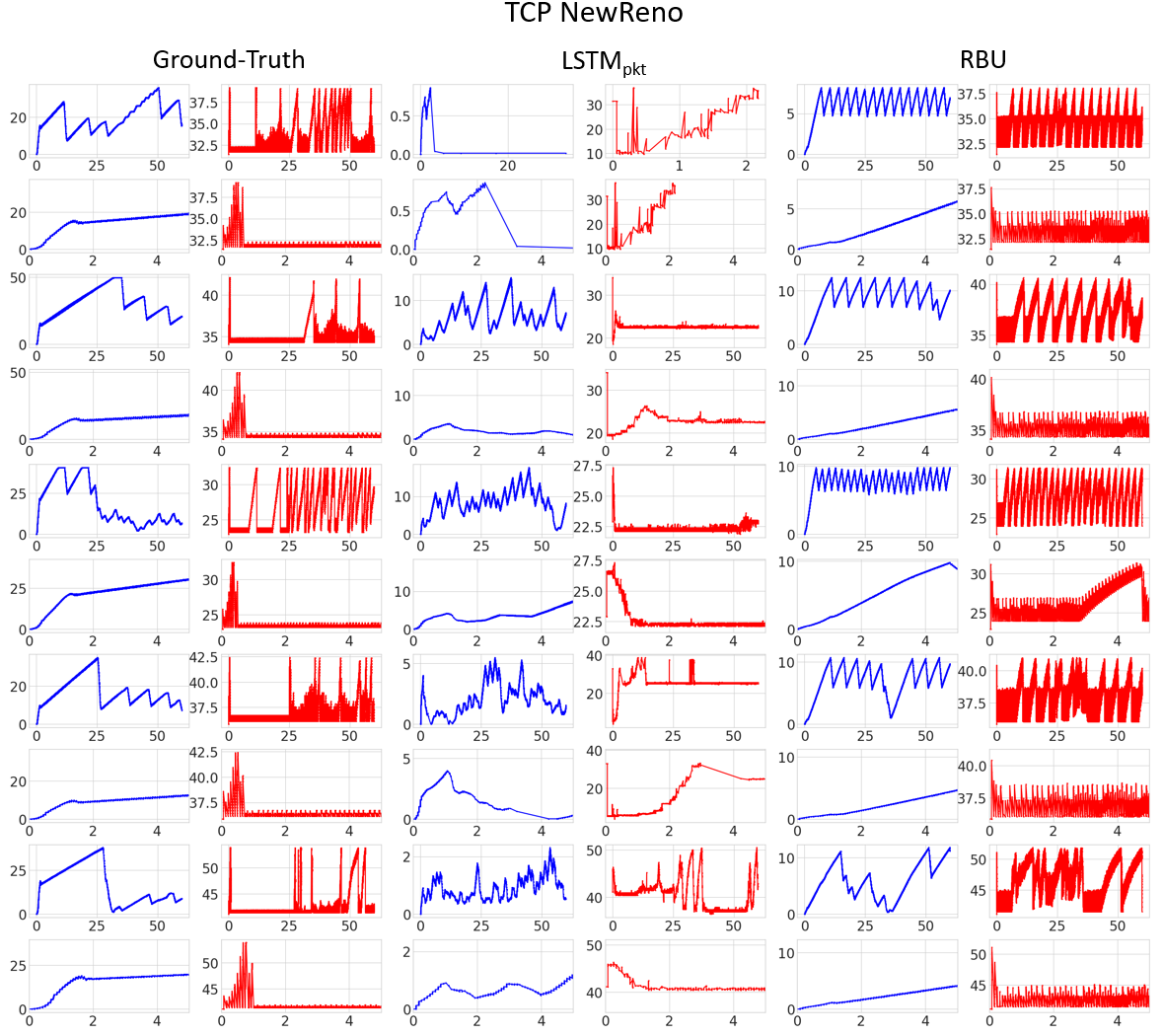}
    \caption{Traces produced by~\RNNSampled~(middle) and the~\buffer~(right) models along with Ground-Truth traces (left) for the test protocol TCP NewReno. The blue curves represent sending rates decided by the test protocol and the red curves represent delays generated by the models. For every pair of rows, the first row shows the entire trace and the second row focuses on the first few seconds. These are randomly picked traces, and there is no one-to-one correspondence between columns.}
    \label{fig:sample-traces-newreno}
\end{figure*}

We present additional example traces for~\RNNSampled~and the~\buffer~models, along with ground truth traces for different protocols in Figures~\ref{fig:sample-traces-cubic},~\ref{fig:sample-traces-vegas},~\ref{fig:sample-traces-ledbat}, and~\ref{fig:sample-traces-newreno}.

\end{document}